# Does Confinement Force Marriage between Two Unwilling Partners?: A Case Study of He$_2$@B$_m$N$_m$ (m=12, 16)

Munmun Khatua, Sudip Pan and Pratim K. Chattaraj*

*Department of Chemistry and Center for Theoretical Studies,*

*Indian Institute of Technology, Kharagpur 721302, India.*

[*]*To whom correspondence should be addressed. E-mail: pkc@chem.iitkgp.ernet.in,*

*Telephone: +91 3222 283304, Fax: 91-3222-255303*

## Abstract

The stability of Ng$_n$@B$_{12}$N$_{12}$ and Ng$_n$@B$_{16}$N$_{16}$ systems is assessed through a density functional study and ab initio simulation. Although they are found to be thermodynamically unstable with respect to the dissociation of individual Ng atoms and parent cage, ab initio simulation reveals that except Ne$_2$@B$_{12}$N$_{12}$ they are kinetically stable to retain their structures intact throughout the simulation time (500 fs). The Ne$_2$@B$_{12}$N$_{12}$ cage dissociates and the Ne atoms get separated as the simulation proceeds. The He-He unit undergoes translation, rotation and vibration inside the cavity of B$_{12}$N$_{12}$ and B$_{16}$N$_{16}$ cages. Electron density analysis (Atoms-in-Molecule (AIM)) shows that there is some degree of covalent character in He-He bond (W$^c$ type) of He$_2$@B$_{12}$N$_{12}$. In case of He$_2$@B$_{16}$N$_{16}$, the He-He interaction is mostly of noncovalent type (W$^n$). In many cases, especially for the heavier Ng atoms the Ng-N/B bonds are found to be of covalent type or at least having some degree of covalent character. But Wiberg bond indices show zero bond order in He-He bond and very low bond order in cases of Ng-N/B bonds. The Ng atoms experience more repulsion in smaller B$_{12}$N$_{12}$ cage than B$_{16}$N$_{16}$ cage. Electron transfer takes place between Ng atoms and cage atoms. However, the amount of electron transfer is low for the lighter Ng atoms which gradually increases for the heavier Ng atoms. The cages undergo deformation in accommodating Ng atoms. Energy decomposition analysis (EDA) provides further insights into the bonding among the Ng atoms and the cage. The overall repulsive nature in interaction energy originates from very high Pauli repulsion ($\Delta E_{pauli}$) although other terms viz., electrostatic interaction ($\Delta E_{elstat}$), orbital interaction ($\Delta E_{orb}$) and the dispersion correction ($\Delta E_{disp}$) are attractive in nature. The change in charge distribution, radius, hardness, electrophilicity and polarizability implies that they possess different properties and reactivity from their parent moieties. The HOMO energies of He$_2$ units in He$_2$@B$_{12}$N$_{12}$ and He$_2$@B$_{16}$N$_{16}$ are significantly higher than that in free He atoms. Their kinetic stability and different calculated properties imply that the He-He interaction and Ng-N/B interaction may be considered as chemical bonds according to the IUPAC definition.

## 1. Introduction

Noble gas chemistry became important since 1962 after the synthesis of Xe$^+$[PtF$_6$]$^-$ by Bartlett[1]. Thereafter, the interest in new noble gas (Ng) compounds took an upsurge owing to the radical departure of a long-established chemical principle, viz., the Ng atoms are inert and do not take part in chemical bonding. Apart from the successful syntheses of a large number of compounds containing heavier Ng atoms[2], important insights into the nature of bonding were obtained using quantum chemical techniques[3]. Before the synthesis of the first Ng compound, Pauling[4] predicted the possibility of chemical bonding for heavier Ng atoms due to their relatively low ionization potential and high polarizability. The contributions of Räsänen and co-workers[5,6] in successfully synthesizing a series of compounds of type H(Ng)Y (Ng=Ar, Kr, Xe; Y=electron-withdrawing group) and Feldman et al.[7] in preparing Ng hydrides and other related Ng compounds are considered as significant in enriching this field. The theoretical prediction of various Ng-compounds by different research workers[8-13] is also equally important for the advancement in this field. The stability of van der Waals complexes of Ng has been studied[14] extensively both computationally and experimentally.

With the advent of fullerene cages, studies involving endohedral confinement of Ng atoms, Ng$_n$@C$_{60}$ and Ng$_n$@C$_{70}$, have been performed both theoretically and experimentally[15]. Ng atoms have been successfully incorporated in the fullerene cages with the use of molecular beams[16]. Effect of high pressure and temperature on the incorporation of Ng into the fullerenes has also been studied[17] and Ng confined in the fullerene cages have been detected in a modified mass spectrometer by heating samples. Experimental studies regarding high-temperature decomposition of Ng@C$_{60}$ complexes showed an activation barrier of 90 kcal/mol[18] during the release of noble gas atoms from $C_{60}$ cage. It has been believed that a cage having an open window made by breaking one or a few bonds is involved as an intermediate in the course of Ng inclusion-exclusion mechanism into the cage[15a,19], although a different opinion is also reported[20]. Ng inclusion into a much smaller cavity than in $C_{60}$ has also been studied[21]. He atom inside the dodecahedrane, $C_{20}H_{20}$ cage has been experimentally identified by Cross et al.[21b], although theoretical calculation reveals that He@$C_{20}H_{20}$ is less stable by 33.8 kcal/mol with respect to the isolated $C_{20}H_{20}$ cage and He atom[21c]. In spite of the repulsive nature of the interaction between He and $C_{20}H_{20}$ cage, it is formed due to its kinetic stability. Once it is formed, it cannot dissociate due to high energy barrier. Therefore, this observation further opens up the hunt for other viable Ng trapped small cages.

Further, the quantum-chemical calculations of Ng dimers ($Ng_2$) (Ng=He-Xe) confined in $C_{60}$ cage[22] have been carried out using DFT and ab initio methods. It has been reported[22a] that the Ng-Ng distances in $Ng_2$@$C_{60}$ are shorter than those in free Ng dimers and these compounds are thermodynamically unstable towards the release of Ng atoms. Krapp and Frenking[22a] declared the existence of a 'genuine' Xe-Xe chemical bond. However for lighter analogues (He and Ne) there are only very weak interactions between them. It is interesting to note that endohedrally confined $Ng_2$@$C_{60}$ systems show a variety of interatomic interactions, hence questioning the "classical" view of a chemical bond. Theoretical studies involving the photoionization of endohedrally confined Ng atoms in $C_{60}$ has also been done[23]. Further, the study of $He_2$@$C_{20}H_{20}$ concerning the influence of endohedral confinement on the electronic interaction between two He atoms has been done by Cerpa et al[24]. Owing to the short internuclear separation of two He atoms in $He_2$@$C_{20}H_{20}$ compared to free He-He distance in $He_2$, it is analyzed that such internuclear separation does not always imply the existence of a chemical bond.

In the present work, in order to study the viability of other smaller cages (than $C_{60}$) to encapsulate Ng atoms and its dimer, we have considered $B_{12}N_{12}$ and $B_{16}N_{16}$ called magic BN-fullerenes[25] for our study. Being isoelectronic with carbon fullerenes, boron nitride cages $(BN)_n$ have been explored widely for both the experimental and theoretical[25a,26] aspects. Density functional studies regarding the endohedrally confined ions/atoms in $B_{16}N_{16}$ nanocage (M@$B_{16}N_{16}$, M = $Li^+$, $Na^+$, $K^+$, $Mg^{2+}$, Ne, $O^{2-}$, $S^{2-}$, $F^-$, $Cl^-$) have been reported earlier[27]. Here, we have performed both electronic structure calculation as well as ab initio molecular dynamics simulation to provide insights into the thermodynamic and kinetic stabilities of Ng encapsulated

$B_{12}N_{12}$ and $B_{16}N_{16}$ cages. Further, we have tried to analyze the probable existence of a chemical bond between Ng-Ng and Ng-cage atoms.

## 2. Theoretical Background

The stability of molecular systems can be understood by assessing the values of hardness ($\eta$) and electrophilicity ($\omega$) through the electronic structure principles like the maximum hardness principle (MHP)[28], minimum polarizability principle (MPP)[29] and minimum electrophilicity principle (MEP)[30]. The electronegativity ($\chi$)[31], hardness ($\eta$)[32] and electrophilicity ($\omega$)[33] for an N-electron system can be expressed as,

$$\chi = -\left(\frac{\partial E}{\partial N}\right)_{v(\vec{r})} = -\mu \quad (1)$$

$$\eta = \left(\frac{\partial^2 E}{\partial N^2}\right)_{v(\vec{r})} \quad (2)$$

and

$$\omega = \frac{\mu^2}{2\eta} = \frac{\chi^2}{2\eta} \quad (3)$$

Here $E$, $\mu$ and $v(\vec{r})$ are the total energy, chemical potential and external potential of the system, respectively. Electronegativity and hardness can be expressed in terms of ionization potential ($I$) and electron affinity ($A$) or the related frontier molecular orbital energies using Koopmans' theorem[34]. Therefore,

$$\chi = \frac{I + A}{2} \quad (4)$$

and

$$\eta = I - A \quad (5)$$

## 3. Computational Details

All the systems considered here have been designed using graphical software GaussView 5.0.8[35]. The $Ng_n@B_{12}N_{12}$ and $Ng_n@B_{16}N_{16}$ systems including the bare $B_{12}N_{12}$ and $B_{16}N_{16}$ cages have been optimized at M05-2X/6-311+G(d,p) level for Ng = He, Ne, Ar, Kr and at M05-2X/def2-TZVP level for Ng = Xe using Gaussian 09 program package[36]. An effective core potential has been used for the core electrons of Xe[37]. Harmonic vibrational frequency analysis has been done for all the structures at the aforementioned levels. Zero NIMAG (number of imaginary frequency) value in all the cases confirms that the structures correspond to minima on the potential energy surface. Natural population analysis (NPA) has been performed to compute atomic charges and Wiberg bond index (WBI)[38] has also been calculated to assess the bond order. The basis set superposition error (BSSE) correction has been done using the counterpoise (CP) method suggested by Boys and Bernardi[39]. To know the nature of Ng-Ng and Ng-N/B interaction in $Ng_n@B_{12}N_{12}$ and $Ng_n@B_{16}N_{16}$ systems, energy decomposition analysis (EDA)[40] and electron density analysis[41] have been made. Contour maps of the Laplacian of the electron density of the Ng-Ng interaction have been plotted using Multiwfn software[42].

The dynamics of all the $Ng_n@B_{12}N_{12}$ and $Ng_n@B_{16}N_{16}$ systems have been investigated by using ab initio molecular dynamics[43], atom-centered density matrix propagation (ADMP)[44] technique as included in Gaussian 09 suite of program[36]. The dynamics have been performed at DFT-D-B3LYP/6-311+G(d,p) level with the above mentioned optimized geometries. Boltzmann distribution has been used to generate the initial nuclear kinetic energies of the systems. The temperature has been maintained by using a velocity scaling thermostat throughout the simulation. Default random number generator seed has been used, implemented in Gaussian 09 to initiate the initial mass weighted Cartesian velocity. For all the cases, trajectories up to 500 fs have been generated. Koopmans' theorem[34] has been used to calculate I and A required to have hardness and electrophilicity.

Dissociation value (D) has been calculated by using the equation (6)

$$D = [(E_{B_{12}N_{12}/B_{16}N_{16}} + nE_{Ng}) - E_{Ng_n@B_{12}N_{12}/B_{16}N_{16}}] \tag{6}$$

## 4. Results and Discussion
### 4.1. Ng@$B_{12}N_{12}$ and Ng@$B_{16}N_{16}$ systems

The optimized geometries of Ng@$B_{12}N_{12}$ and Ng@$B_{16}N_{16}$ are presented in Figures 1 and 2, respectively, and Table 1 shows the dissociation energy ($D_e$), BSSE corrected dissociation energy ($D_{BSSE}$), zero point energy (ZPE) corrected dissociation energy ($D_0$), both ZPE and BSSE corrected dissociation energy ($D_0^{BSSE}$), reaction enthalpy (ΔH), free energy change (ΔG), hardness (η), electrophilicity (ω) and polarizability (α) for the bare as well as Ng encapsulated $B_{12}N_{12}$ and $B_{16}N_{16}$ cages. The negative values of dissociation energies and positive values of reaction enthalpies and free energy changes indicate that the Ng entrapping processes into the

$B_{12}N_{12}$ and $B_{16}N_{16}$ cages are thermodynamically unfavorable. The $D_{BSSE}$ values are slightly higher[45] than the corresponding $D_e$ values for all the systems; whereas in case of $D_0$ values, they are higher than the corresponding $D_e$ values for He and Ne cases but for the rest of the systems they turn out to be smaller than those of $D_e$ values. It shows that for He and Ne cases, the correction from ZPE in Ng trapped cages is larger than those in the empty cages but for the heavier Ng entrapped analogues, the reverse is true. Here we have discussed only the both ZPE and BSSE corrected dissociation energy values ($D_0^{BSSE}$).

In Ng@$B_{12}N_{12}$ systems, the $D_0^{BSSE}$ values are -34.9 kcal/mol for He, -97.6 kcal/mol for Ne, -305.2 kcal/mol for Ar, -431.9 kcal/mol for Kr and -620.1 kcal/mol for Xe cases, respectively, whereas in Ng@$B_{16}N_{16}$ systems, the $D_0^{BSSE}$ values are -9.2 kcal/mol for He, -27.0 kcal/mol for Ne, -115.0 kcal/mol for Ar, -181.6 kcal/mol for Kr and -293.4 kcal/mol for Xe cases, respectively. From the dissociation values it is observed that the destabilization caused by the Ng encapsulation into $B_{16}N_{16}$ cage is noticeably smaller than that in the case of $B_{12}N_{12}$ cage and with an increase in the size of the Ng atoms, the encapsulation process becomes gradually unfavorable. The first one is due to the larger space available to the Ng atoms in the $B_{16}N_{16}$ cage. It has already been established[22a,24,46] that for such type of Ng entrapped cages, the destabilization originates from the cage distortion ($E_{prep}$, the required energy to distort the cage in Ng@Cage) and Pauli repulsion ($\Delta E_{Pauli}$). Therefore, with an increase in the cage cavity, to afford an Ng atom inside the cage, the necessity of cage distortion and the possibility of Pauli repulsion originated from the repulsion of two electrons having same spin diminish. This is the reason why Ng atoms interact with larger $C_{60}$ cage with attractive interaction[15b] but the same for smaller $C_{20}H_{20}$ cage is repulsive in nature[21b]. On the other hand, with an increase in Ng size, both the degree of cage distortion and Pauli repulsion will increase. The $\Delta H$ and the $\Delta G$ values also reflect the same trend as set by the dissociation energy values i.e., they become more and more positive with an increase in the size of the Ng atoms and for a particular Ng atom, the corresponding values are less for $B_{16}N_{16}$ than those for $B_{12}N_{12}$. Note that in experimentally synthesized He@$C_{20}H_{20}$, the dissociation energy is -33.8 kcal/mol (-98.3 kcal/mol for Ne in Ne@$C_{20}H_{20}$)[21a,21b]. In comparison to these two values, $D_0^{BSSE}$ values for He and Ne trapping in $B_{16}N_{16}$ cage are lesser, therefore, more likely to be formed. The computed maximum B-B and N-N distances show that in going from He to Xe, the cage is gradually more expanded i.e., undergoes distortion (see Table 2).

The values of the global reactivity descriptors like η, ω and α presented in Table 1 show that with the encapsulation of Ng atoms into the $B_{12}N_{12}$ cage, η decreases and ω increases (except the case of Xe) and α increases for all cases with respect to the bare cage. Hence, with the aid of MHP, MEP and MPP, it can be inferred that the endohedral confinement of Ng atoms by the $B_{12}N_{12}$ cage renders a loss of electronic stability of this system. However, in the case of Ng encapsulated $B_{16}N_{16}$ cages, the hardness and electrophilicity values in Ng@$B_{16}N_{16}$ increase and decrease, respectively, with respect to $B_{16}N_{16}$ (except Xe@$B_{16}N_{16}$). Therefore, although the dissociation energy values imply the repulsive type of interaction between Ng and cage, η and ω values show higher electronic stability of Ng trapped systems than the empty one. Nevertheless, α values of Ng entrapped analogues are higher than $B_{16}N_{16}$ cage implying increasing softness. Note that as the size of the Ng atoms increases, η decreases while ω and α increase.

### 4.2. Ng$_2$@$B_{12}N_{12}$ and Ng$_2$@$B_{16}N_{16}$ systems

Fascinated by the earlier studies on Ng$_2$@$C_{60}$[22a] and He$_2$@$C_{20}H_{20}$[24], to get further insight into the nature of bonding and stability, here we have considered the encapsulation of two Ng atoms into the $B_{12}N_{12}$ and $B_{16}N_{16}$ cages. Ng$_2$@$B_{12}N_{12}$ (Ng=He, Ne) and He$_2$@$B_{16}N_{16}$[47] turn out as minimum energy structures (see Figure 3). Upon free optimization, the Ng dimers in $B_{12}N_{12}$ and $B_{16}N_{16}$ cages orient towards the mid point of the six membered rings composed by B and N atoms. We do not perform any pre-assignment of symmetry on the basis of the previous studies [22a,24]. Due to the free Ng$_2$ precession, they concluded that symmetry assignment is not pertinent. Here we have also observed similar type of Ng$_2$ precession in our ab initio simulation, which will be discussed in the next section. Similar to Ng$_2$@$C_{60}$[22a] and He$_2$@$C_{20}H_{20}$[24], in Ng$_2$@$B_{12}N_{12}$ (Ng=He, Ne) and He$_2$@$B_{16}N_{16}$ systems the Ng-Ng bonds are also found to be considerably smaller than their free dimers. In He$_2$@$B_{12}N_{12}$ and He$_2$@$B_{16}N_{16}$, the He-He distances are 1.306 Å and 1.456 Å, respectively, whereas the He-He bond length in free He$_2$ dimer is 2.666 Å. Likewise for the Ne$_2$@$B_{12}N_{12}$ system, the Ne-Ne length (1.608 Å) is shorter than that in the free Ne$_2$ dimer (2.899 Å). If we consider the dissociation of Ng$_2$@cage into two free Ng atoms and vacant cage, $D_0^{BSSE}$ values (Table 1) are -163.1 kcal/mol (-81.6 kcal/mol per He atom) in He$_2$@$B_{12}N_{12}$, -455.6 kcal/mol (-227.8 kcal/mol per Ne atom) in Ne$_2$@$B_{12}N_{12}$, and -72.9 kcal/mol (-36.5 kcal/mol per He atom) in He$_2$@$B_{16}N_{16}$ systems. The $D_e$ and $D_{BSSE}$ values for He$_2$@$B_{12}N_{12}$ are -129.2 kcal/mole and -130.3 kcal/mole, respectively, and for He$_2$@$B_{16}N_{16}$ are -52.6 kcal/mole and -53.6 kcal/mole, respectively, when we consider the dissociation of He$_2$@$B_{12}N_{12}$

and He$_2$@B$_{16}$N$_{16}$ cages into the bare B$_{12}$N$_{12}$ and B$_{16}$N$_{16}$ cages and free He$_2$ dimer having He-He distance same as in He$_2$@B$_{12}$N$_{12}$ and He$_2$@B$_{16}$N$_{16}$. In vacant B$_{12}$N$_{12}$ and B$_{16}$N$_{16}$ cages, B-B$^{Max}$/N-N$^{Max}$ distances are 4.368/4.808 Å and 4.908/5.367 Å, respectively, but in He$_2$@B$_{12}$N$_{12}$ and He$_2$@B$_{16}$N$_{16}$ cages, they elongate to some extent and become 4.554/4.961 Å and 5.022/5.400 Å, respectively (see Table 2). It shows that the cages get distorted in accommodating two He atoms. Note that in He@B$_{12}$N$_{12}$ and He@B$_{16}$N$_{16}$ cages, the distortion is quite small. The relevant molecular orbitals (MOs), which involve in the He-He interaction, are provided in the Figure S1 (in supporting information). The HOMO energies of He$_2$ units having bond lengths as those in He$_2$@B$_{12}$N$_{12}$ and He$_2$@B$_{16}$N$_{16}$ have been found to increase by 2.954 eV and 2.094 eV, respectively, compared to the highest occupied atomic orbital (HOAO) energy of free He atom (see Table 3). The difference in HOMO energy of [He$_2$] and that MO energy corresponding to the He-He interaction in He$_2$@B$_{12}$N$_{12}$ show a destabilization (by 0.119 eV) of the valence electrons of He$_2$ in the cage.

### 4.3. Ab initio simulation

To understand the stability and dynamical behavior, an ab initio molecular dynamics calculation has been performed. Time evolution of energies of Ng$_n$@B$_{12}$N$_{12}$ and Ng$_n$@B$_{16}$N$_{16}$ systems is presented in Figures 4 and 5, respectively. The snapshots of the structures at t = 0 and 500 fs for Ng@B$_{12}$N$_{12}$ and Ng@B$_{16}$N$_{16}$ systems are presented in Figures S2 and S3 (in supporting information). Oscillations in energy of Ng@B$_{12}$N$_{12}$ systems are observed except for Ne$_2$@B$_{12}$N$_{12}$ system. This may be due to the increase in nuclear kinetic energy resulting in the distortion of the cages. The confined Ng atoms float within the cage throughout the simulation and when they approach the walls of the cage, distortions occur, producing oscillations in the energy curves. For smaller Ng atoms, the movement is more noticeable than the larger Ng atoms. This is due to the larger space available for their movement. For all the systems except for Ne$_2$@B$_{12}$N$_{12}$, the energy oscillates within a fixed range. The energy of Ne$_2$@B$_{12}$N$_{12}$ decreases with time; the reason behind such decrease in energy can be understood from Figure 6. At the end of the simulation i.e., at t = 500 fs, we can see that both the Ne atoms are separated from the B$_{12}$N$_{12}$ cage. It is interesting to note that at t = 12 fs, one of the hexagonal faces of the B$_{12}$N$_{12}$ cage gets distorted as one of the Ne atoms approaches that face and at t = 60 fs the distortion increases, which is reflected in the increase in energy at these two time steps (Figure 4). At t =

200 fs and 335 fs, the first and the second Ne atoms get out of the cage, respectively, producing a local minimum on the energy surface. This observation further supports the proposed 'window' mechanism in Ng inclusion-exclusion process into the cage1[5a,18,19]. At t = 460 fs, the cage retracts back to its initial shape resulting in the final picture as presented in Figure 6. As the endohedral confinement of both the Ne atoms is thermodynamically unstable towards the release of the atoms, once the cage opens up both the atoms move out of the $B_{12}N_{12}$ cage. Although the endohedral confinement of Ng atoms is thermodynamically unstable towards the release of Ng atoms, all the structures remain intact throughout the simulation process (500fs) except of $Ne_2@B_{12}N_{12}$. Hence, we can conclude that such structures are stable enough to exist at least kinetically if not thermodynamically.

Till now, Straka et al.,[22a] Frenking et al.,[22b] and Merino et al.[24] referred to the free precession of $He_2$ inside the $C_{60}$ or $C_{20}H_{20}$ cage on the basis of very low energy difference between the isomers having different He-He orientation and/or on the basis of $^3$He NMR data. Here, through the ab initio simulation we are first time demonstrating the precession, encompassing translation, rotation and vibration, of $He_2$ unit inside the $B_{12}N_{12}$ and $B_{16}N_{16}$ cavity with time. Due to larger space the precession of $He_2$ is much clearer in $B_{16}N_{16}$ than that in $B_{12}N_{12}$. In Figures 7 and 8, we have depicted the snapshots of $He_2@B_{12}N_{12}$ and $He_2@B_{16}N_{16}$ systems at different time steps, respectively, to show the precession of $He_2$ unit.

Time evolution of hardness and electrophilicity of $Ng_n@B_{12}N_{12}$ and $Ng_n@B_{16}N_{16}$ (Ng=He-Kr) systems are presented in Figure 9, whereas those for $Xe@B_{12}N_{12}$ and $Xe@B_{16}N_{16}$ systems are given in Figure 10. The hardness and electrophilicity curves show presence of oscillations throughout the simulation similar to those of the energy curves. Only in case of $Ne_2@B_{12}N_{12}$ system, hardness and electrophilicity values increase and decrease, respectively. This is due to the favorable dissociation of $Ne_2@B_{12}N_{12}$ into $B_{12}N_{12}$ and Ne atoms. For other $Ng_n@B_{12}N_{12}$ systems, both remain almost unchanged throughout the simulation. Comparing the hardness and electrophilicity of the $Ng@B_{12}N_{12}$ (Ng=He-Kr) systems in Figure 9, it is clear that hardness decreases and electrophilicity increases from He to Kr i.e., stability of the systems decrease in the same order.

**4.5. Nature of interaction**

In this section, we have repeated the same question; 'Is this a chemical bond?' as Frenking et al.[22a] did in the case of $Ng_2@C_{60}$. To know the answer whether confinement can induce a bond between two unwilling atoms (He-He or Ng-B/N), we have performed Wiberg bond index (WBI) calculation, natural population analysis (NPA), the detailed electron density analysis (AIM). We have also highlighted the outcome of our ab initio simulation and associated change in physical properties to draw a possible conclusion in this regard.

**4.5.1. Natural population analysis and Wiberg bond indices**

NPA calculation shows that for all the systems, electron transfer takes place from the Ng atoms to the cage atoms; hence positive charge develops on the Ng atoms. The ionization energies of Ng atoms are high, being the highest for the He atom (compared to other Ng atoms) and the electron affinities of the $B_{12}N_{12}$ (0.028 eV) and $B_{16}N_{16}$ (0.589 eV) cages are low. So it is expected that any electronic transfer from the Ng atoms to the cage will be very small, being smallest for the He systems. Indeed the net NPA charges on the Ng atoms in the $B_{12}N_{12}$ cages are +0.089 $e^-$ (on He atom in $He@B_{12}N_{12}$), +0.085 $e^-$ (on each He atoms in $He_2@B_{12}N_{12}$), +0.107 $e^-$ (on Ne atom in $Ne@B_{12}N_{12}$), +0.121 $e^-$ (on each Ne atoms in $Ne_2@B_{12}N_{12}$), +0.317 $e^-$ (on Ar atom in $Ar@B_{12}N_{12}$), +0.486 $e^-$ (on Kr atom in $Kr@B_{12}N_{12}$) and +0.775 $e^-$ (on Xe atom in $Xe@B_{12}N_{12}$) (see Table 2). Therefore, the shifting of the electron from the Ng atoms to the cage atoms increases from He to Xe as the ionization energies decrease along the same order. Note that in $Ng_n@B_{12}N_{12}$ and $Ng_n@B_{16}N_{16}$ systems the N and B centers acquire more negative and positive charges, respectively, than those in vacant cages (Table 2). The total WBI for He atom in $He@B_{12}N_{12}$ is 0.175, for each He atoms in $He_2@B_{12}N_{12}$ is 0.174, for Ne atom in $Ne@B_{12}N_{12}$ is 0.226, for each Ne atoms in $Ne_2@B_{12}N_{12}$ is 0.252, for Ar atom in $Ar@B_{12}N_{12}$ is 0.752, for Kr atom in $Kr@B_{12}N_{12}$ is 1.125 and for Xe atom in $Xe@B_{12}N_{12}$ is 1.607. The individual Ng-N/B bonds show a low value of WBI in the order of $10^{-2}$ in all cases. Low WBIs suggest the presence of van der Waals interaction between the Ng atoms and the cage atoms. As the size of the Ng atom increases, the total WBI value increases. The gradual increase in the total WBI values shows the increase in interaction between the Ng atoms and the cage atoms. In case of $He_2@B_{12}N_{12}$, although there is a reduction in the He-He length, the WBI value shows a zero bond order. Therefore, WBI value indicates that the decrease in the He-He distance is due to the endohedral confinement rather than the He-He bond formation. Now, we have considered the

$Ng_n@B_{16}N_{16}$ systems. The net NPA charges on the Ng atoms in the $B_{16}N_{16}$ cages are + 0.058 e⁻ (on He atom in $He@B_{16}N_{16}$), + 0.066 e⁻ and + 0.067 e⁻ (on two He atoms in $He_2@B_{16}N_{16}$), + 0.080 e⁻ (on Ne atom in $Ne@B_{16}N_{16}$), + 0.256 e⁻ (on Ar atom in $Ar@B_{16}N_{16}$), + 0.377 e⁻ (on Kr atom in $Kr@B_{16}N_{16}$) and + 0.458 e⁻ (on Xe atom in $Xe@B_{16}N_{16}$) (see Table 2). Similar to the $Ng_n@B_{12}N_{12}$ systems, here also the positive charges on the Ng atoms increase from He to Xe. Although the net charges on the Ng atoms for the $Ng_n@B_{16}N_{16}$ systems are lower as compared to those of $Ng_n@B_{12}N_{12}$ systems. This can be due to the larger space available in the $B_{16}N_{16}$ cage, which reduces the possibility of electronic transfer. The total WBI for He atom in $He@B_{16}N_{16}$ is 0.117, for each of the two He atoms in $He_2@B_{16}N_{16}$ are 0.135 & 0.136, for Ne atom in $Ne@B_{16}N_{16}$ is 0.166, for Ar atom in $Ar@B_{16}N_{16}$ is 0.567, for Kr atom in $Kr@B_{16}N_{16}$ is 0.821 and for Xe atom in $Xe@B_{16}N_{16}$ is 0.929. The WBI values for the $Ng@B_{16}N_{16}$ systems are lower as compared to those in $Ng@B_{12}N_{12}$ systems. In $He_2@B_{16}N_{16}$ also, the WBI value shows a zero bond order between two He atoms. In all cases, the WBIs for Ng-N/B bonds are in the order of $10^{-3}$ to $10^{-2}$. The valence orbital population of the Ng atoms for all the systems is presented in Table 4. It gives us an idea about the orbitals taking part in electron transfer from Ng atoms to the cage atoms. The valence orbital population of Ng atoms shows that the electron transfer takes place from the s orbital and all the three p orbitals. Also from Ne-Kr the s orbital contributes more to the electron transfer than the p orbitals. But for the case of Xe, the p orbitals contribute more.

### 4.5.2. Electron density analysis

More insight into the He-He bonding nature in the $He_2@B_{12}N_{12}$ and $He_2@B_{16}N_{16}$ systems can be obtained from the analysis of the electron density[41]. The topological parameters obtained at the bond critical point (BCP) between He-He bonds or Ng-N/B bonds from the atoms in molecules (AIM) analysis are provided in Table 5. A negative value of Laplacian of electron density ($\nabla^2\rho(r_c)$) at the BCP implies electron density concentration whereas positive value indicates electron density depletion. Hence, generally negative and positive values of $\nabla^2\rho(r_c)$ represent the covalent and non covalent interaction, respectively. But this condition does not seem sufficient to describe the systems with 3d or heavier atoms[48]. Even it cannot describe some typical covalent molecules (like $F_2$, CO etc.)[41;pp 312-314]. Other parameters like local kinetic energy density ($G(r_c)$), local potential energy density ($V(r_c)$), local electron energy density ($H(r_c)$), and ratio of $-G(r_c)/V(r_c)$ and $G(r_c)/\rho(r_c)$ have also been calculated to get further insight into this. $H(r_c)$ is calculated using the equation given below.

$$H(r_C) = G(r_C) + V(r_C) \qquad (7)$$

Cremer et al.[49] suggested that if $\nabla^2\rho(r_c) > 0$ and $H(r_c) < 0$, then the bonding is partly of covalent type. Further, the value of $-G(r_c)/V(r_c)$ having greater than 1 indicates a purely noncovalent type of interaction whereas if it falls in between 0.5 and 1 then there exists some degree of covalent character (partial covalent character)[50]. The ratio of $G(r_c)/\rho(r_c)$ has also been employed as an indicator of covalent bond[48]. Generally the value of $G(r_c)/\rho(r_c)$ less than 1 indicates the presence of covalent bonding. Boggs et al.[51] have performed very recently the AIM analysis of a series of systems having Ng atoms. They have classified the covalent bonds into four types[51].

Type A. $\nabla^2\rho(r_c) < 0$, and $\rho(r_c)$ is large ($\geq 0.1$ au)

Type B. $H(r_c) < 0$, and $\rho(r_c)$ is large ($\geq 0.1$ au)

Type C. $H(r_c) < 0$, and $G(r_c)/\rho(r_c) < 1$

Type D. $|H(r_c)|$ is small ($< 0.005$ au) and $G(r_c)/\rho(r_c) < 1$

They have also proposed two new categories viz., $W^c$ (weak interaction having some degree of covalent character) and $W^n$ (weak interaction having noncovalent character). In our case, we have also followed these categories to assign the type of bonding (see Table 5). The molecular graphs generated by AIM2000 program[52] and provided in Figure S4 (supporting information) show that there is always a bond path between two He atoms in $He_2@B_{12}N_{12}$ and $He_2@B_{16}N_{16}$ and of course, existence of a bond path between two atoms needs not imply the existence of a chemical bond, rather it tells that they are bonded[53]. For the lighter Ng analogues, mostly bond paths between Ng and N atoms have been observed, whereas in heavier analogues, bond paths between Ng and both N and B atoms have been obtained (see Figure S4). In $He_2@B_{12}N_{12}$ and $He_2@B_{16}N_{16}$, each He atom is connected to the three N atoms via bond path. The numbers of bond paths are 6 in $He@B_{12}N_{12}$, 12 in $Ne@B_{12}N_{12}$, 12 in $Ar@B_{12}N_{12}$, and 24 in $Kr@B_{12}N_{12}$, whereas they are 5 in $He@B_{16}N_{16}$, 14 in $Ne@B_{16}N_{16}$, 14 in $Ar@B_{16}N_{16}$, and 20 in $Kr@B_{16}N_{16}$. Now let us look at Table 5 to see in which category these bonds fall. In case of $Ng@B_{12}N_{12}$, for Ng=He-Ar, the Ng-N interaction falls in the type of $W^n$, whereas for Kr analogue, the Kr-N bonds fall in the type of $W^c$ and Kr-B bonds fall in the type of C. In case of $Xe@B_{12}N_{12}$, both

Xe-N and Xe-B bonds are of C type. In He@$B_{16}N_{16}$, both He-N and He-B bonds are of D type. In Ne@$B_{16}N_{16}$ and Ar@$B_{16}N_{16}$, except Ar-B bonds (C type), all Ng-N/B bonds are of $W^n$ type. All the Ng-N/B bonds are of C type except Kr-N bonds (D type) in Kr@$B_{16}N_{16}$ and Xe@$B_{16}N_{16}$. We are particularly interested on the nature of He-He bonds in $He_2$@$B_{12}N_{12}$ and $He_2$@$B_{16}N_{16}$. The He-He bond in $He_2$@$B_{12}N_{12}$ falls under the category $W^c$ type, whereas that in $He_2$@$B_{16}N_{16}$ falls in the borderline of $W^c$ and $W^n$ with $H(r_c)$ value of 0.0 au and -$G(r_c)$/$V(r_c)$ value of 1.0 au. Therefore, this result shows that with the confinement through a small cage, one can induce at least some degree of covalent character between two He atoms. We have also considered NgBe$CN_2$[9f] and NgBeO[8d] (Ng=Kr,Xe) in our present study to compare these result with that of He-He bond. It has been believed that for the heavier Ng atoms, some degree of covalent character ($W^c$ type) exists between Be–Ng bonds[9f,51]. The results corresponding to these systems are very similar to that of He-He bond in $He_2$@$B_{12}N_{12}$, which further establishes the existence of some degree of covalent character in He-He bond (see Table 5). Note that in the study of Frenking et al.[22a], they got positive $H(r_c)$ value for He and Ne whereas it turned out negative for Ar-Xe. Therefore, confinement in a smaller cage can improve the bonding situation between two unwilling atoms. Figure 11 presents the contour lines diagram of the Laplacian of electron density, $\nabla^2\rho(r_c)$ of the $He_2$@$B_{12}N_{12}$ and $He_2$@$B_{16}N_{16}$. The negative and positive values of $\nabla^2\rho(r_c)$ have been shown by the dashed blue lines and solid green lines, respectively. It is clear from the figure that there is no charge concentration along the connecting path between the helium atoms. There is only a little deformation in the charge depletion area along the same.

### 4.5.3. Energy decomposition analysis

A detailed account of the energy decomposition analysis (EDA) has been provided in Table 6. At first, the dissociation energy values have been decomposed into two terms viz., interaction energy ($\Delta E_{int}$) and preparation energy ($\Delta E_{prep}$) as –$D_e$ = $\Delta E_{int}$ + $\Delta E_{prep}$. A positive $\Delta E_{int}$ value indicates repulsive interaction between two fragments, whereas a negative one shows the presence of an attractive interaction therein. The $\Delta E_{prep}$ term represents the required energy to distort the cage from its individual equilibrium geometry to adopt Ng atoms inside and to bring two He atoms at a shorter distance as in the complexes. The calculated $\Delta E_{int}$ and $\Delta E_{prep}$ values at M05-2X/6-311+G(d,p) level are given in Table S1 (supporting information). With an increase in the size of Ng atoms, the cages undergo more and more deformation, which is clearly reflected in the $\Delta E_{prep}$ values for the cages. It is low for smaller Ng atoms but for larger Ng atoms it is considerably large. Note that with an increase in the cage size (in case of $B_{16}N_{16}$), the cage undergoes less distortion. Therefore, the corresponding $\Delta E_{prep}$ values are also less than those in the smaller cage ($B_{12}N_{12}$). The $\Delta E_{prep}$ value for $He_2$ dimer in $B_{12}N_{12}$ cage is larger than that in $B_{16}N_{16}$ cage. This is due to the smaller He-He bond distance in $He_2@B_{12}N_{12}$ than that in $He_2@B_{16}N_{16}$. The interaction energy ($\Delta E_{int}$) values are further decomposed into four terms viz., Pauli repulsion ($\Delta E_{pauli}$), electrostatic interaction energy ($\Delta E_{elstat}$), orbital interaction ($\Delta E_{orb}$) and the dispersion correction ($\Delta E_{disp}$) terms and are calculated at BP86-D/TZ2P//M05-2X/6-311+G(d,p) level (see Table 6). Note that in all cases, the $\Delta E_{elstat}$, $\Delta E_{orb}$ and $\Delta E_{disp}$ terms are attractive in nature having $\Delta E_{elstat}$ as the most dominating term among them but the $\Delta E_{pauli}$ term is so much repulsive that it makes the overall interaction energy repulsive in nature. With an increase in the size of Ng atoms, all of $\Delta E_{pauli}$, $\Delta E_{elstat}$, $\Delta E_{orb}$ terms increase but the destabilization provided by the $\Delta E_{pauli}$ increases much more steeply than the other attractive terms. Now, we are particularly interested in the bonding situation in the $He_2$ dimer having shorter bond lengths as in $He_2@B_{12}N_{12}$ and $He_2@B_{16}N_{16}$. It should be noted that a small bond distance in $He_2$ not only increases the destabilizing $\Delta E_{pauli}$ factor, but also induces some degree of attractive $\Delta E_{elstat}$ and $\Delta E_{orb}$ terms, which are -7.8 kcal/mol and -5.4 kcal/mol, respectively, in $He_2$ having the same bond length as in $He_2@B_{12}N_{12}$ and -4.2 kcal/mol and -2.5 kcal/mol, respectively, in $He_2$ having the same bond length as in $He_2@B_{16}N_{16}$. The IUPAC definition of chemical bond is "there is a chemical bond between two atoms or groups of atoms in the case that the forces acting between them are such as to lead to the formation of an aggregation with sufficient stability to make it convenient for the chemist to consider it as an independent 'molecular species'"[54]. Therefore, it is not needed to have an attractive interaction between two fragments to call it a chemical bond rather that should be of sufficient stability to be considered as independent 'molecular species'. In our case, such large $\Delta E_{pauli}$, $\Delta E_{elstat}$, $\Delta E_{orb}$ terms originating from the interaction between Ng atoms and cages will definitely make them independent molecular systems having different property than those of the parent systems.

**4.5.4. Observation from the ab initio simulation**

We want to highlight here the observation coming out from the simulation. In the simulation, one can see the precession of $He_2$ units. In their movement, it appears that they are following each other, no randomness is there. Hence, they are highly correlated to each other. It can be better understood from the plot of the He-He bond distance vs time (see Figure S5). Except the usual stretching and compression of He-He bond, not much abruptness appears in the plot. Therefore, there is definitely some type of bond between two He atoms. It is akin to two unwilling partners forced to live in a very small room. With time, sensing that they have to live together, they made some understanding and learnt to live in a better way by cooperating with each other.

**4.5.5. Arguments based on the proposed physical properties and reactivity**

Based on the IUPAC definition of chemical bond[54], it is worthwhile to look into their properties and reactivity to check whether they may be considered to be independent molecular systems. If they are different from that of the parent moieties, then one may call that there exists a chemical bond. Based on this, Frenking[22a] argued about the existence of chemical bond in $Ar_2@C_{60}$ and $Kr_2@C_{60}$. Note that due to the larger size of $C_{60}$, the electron transfer from Ng to cage atoms is quite small. Ar and Kr acquire +0.01 au and +0.03 au electronic charge in $Ar_2@C_{60}$ and $Kr_2@C_{60}$, respectively[22a]. In our case, due to smaller cage size, the electron shift is quite larger, even in $He@B_{12}N_{12}$ and $He@B_{16}N_{16}$, the NPA charge on He atoms are +0.089 au and +0.058 au, respectively (see Table 2). Additionally, due to smaller sizes than $C_{60}$, here the deformation of the cages due to Ng atoms inclusion will be greater than that of $C_{60}$ case. The NPA charges on each B and N atoms and $r_{B-B}^{Max}$/ $r_{N-N}^{Max}$ distances in Table 2 clearly show the differences in Ng trapped analogues with that of the empty cage. Therefore, these two together definitely alter the reactivity and properties of Ng encapsulated systems from the bare one. The conceptual DFT based reactivity descriptors like η, ω and α in Table 1, clearly reflect that there is a definite change in their values in Ng trapped analogues compared to empty cages. We have found that the HOMOs of $He_2$ in $B_{12}N_{12}$ and $B_{16}N_{16}$ are respectively 2.954 eV and 2.094 eV higher in energy than that of He atom (see Table 3). It shows that the reactivity of He atoms will be different than that of free He atoms. We have further carried out NMR study taking $He_2@B_{12}N_{12}$ and $He_2@B_{16}N_{16}$ as examples (see Figure S6 in supporting information) and found that the magnetic shielding of B and N atoms changes in He trapped analogues compared to the empty one. Therefore, all these observations show that they can be considered as "independent

'molecular species'" having different properties than their parent moieties. Hence, according to the IUPAC convention, we may say that chemical bond exists in these systems.

Therefore, from all the above discussion, one can understand that some indicators like WBI, topological descriptors in some cases point out a closed-shell type of bonding. On the other hand, energy decomposition analysis, the observation from the simulation, the occurrence of different properties and reactivity from that of the parent moiety and the topological descriptors in many cases hint about the presence of chemical bonding therein. One aspect is clear that the bonds involving Ng atoms cannot be explained in terms of conventional model of bonding. As we know, chemical bonding is a fuzzy concept, not a real object, may be it is possible to avoid any debate by categorizing these types of bonds as forming a new class, viz. noble gas type of bond as proposed by Boggs et al.[51]. In this regard, we may recall the remark of Roald Hoffmann regarding the chemical bonding "I think that any rigorous definition of a chemical bond is bound to be impoverishing, leaving one with the comfortable feeling, "yes (no), I have (do not have) a bond," but little else. And yet the concept of a chemical bond, so essential to chemistry and with a venerable history, has life, generating controversy and incredible interest. My advice is this: Push the concept to its limits. Be aware of the different experimental and theoretical measures out there. Accept that at the limits a bond will be a bond by some criteria, may be not others. Respect chemical tradition, relax, and instead of wringing your hands about how terrible it is that this concept cannot be unambiguously defined, have fun with the fuzzy richness of the idea."[55].

5. Conclusion

Both density functional theory study and ab initio simulation have been carried out to access the stability of endohedrally Ng trapped $B_{12}N_{12}$ and $B_{16}N_{16}$ analogues. It has been found that they all are minima on the PES although they are thermodynamically unstable with respect to the dissociation into the related bare cage and Ng atoms. With an increase in the size of the Ng atoms, the repulsive interaction offered by the cage increases. Due to the larger cavity available in $B_{16}N_{16}$ cage compared to $B_{12}N_{12}$ cage, encapsulated Ng atoms face lesser repulsion in $Ng_n@B_{16}N_{16}$ than those in $Ng_n@B_{12}N_{12}$. Similar to those in $Ng_2@C_{60}$[22a] and $He_2@C_{20}H_{20}$[24], a small Ng-Ng bond distance is found in $Ng_2@B_{12}N_{12}$ and $Ng_2@B_{16}N_{16}$. Although WBI calculation shows a zero bond order and very low bond order between He-He and Ng-N/B bonds, respectively, electron density analysis shows that He-He bond in $He_2@B_{12}N_{12}$ is of $W^c$ type, whereas in many cases Ng-N/B bonds are of covalent type (type C and D) or W[c] type. NPA charge calculation reveals that electron transfer takes place from Ng atoms to the cage atoms and it is low for lighter Ng atoms but considerably higher for heavier analogues. The alteration of NPA charge at each center and distortion of cage due to inclusion of Ng atoms causes the change of their properties and reactivity from the parent moieties. EDA analysis reveals that in all cases $\Delta E_{elstat}$, $\Delta E_{orb}$ and $\Delta E_{disp}$ are attractive in nature. $\Delta E_{elstat}$ is the most dominant attractive term among them. $\Delta E_{pauli}$ is so repulsive in nature that overall interaction energy between Ng atoms and cage becomes repulsive. With an increase in the size of Ng atoms, the contributions from all of $\Delta E_{pauli}$, $\Delta E_{elstat}$ and $\Delta E_{orb}$ terms increase. However, the resulting systems become more unstable (with respect to dissociation) due to a larger increase in unfavorable $\Delta E_{pauli}$ over attractive $\Delta E_{elstat}$ and $\Delta E_{orb}$ terms. The corresponding energy terms are found to be smaller in larger $Ng_n@B_{16}N_{16}$ cage than those in the $Ng_n@B_{12}N_{12}$ cage. Further, the change in hardness, electrophilicity and polarizability values from the bare one hints about the change in their properties and reactivity. The He atoms in $He_2@B_{12}N_{12}$ and $He_2@B_{16}N_{16}$ are found to be more reactive than that of free He atoms. Therefore, with the help of IUPAC definition, we may conclude about the existence of chemical bonds in the studied systems. Definitely these bonds cannot be explained on the basis of conventional bonding models. It will be better if we categorize these types of bonds in a different class, viz. noble gas type of bonds, having some interesting features. Ab initio molecular dynamics simulation shows that up to 500 fs, all the systems remain intact except in the case of $Ne_2@B_{12}N_{12}$ in which Ne atoms come out of the cage. Therefore, it shows that although these systems are thermodynamically unstable with respect to dissociation into individual fragments they may be stable kinetically. The He-He bonds in $He_2@B_{12}N_{12}$ and $He_2@B_{16}N_{16}$ are found to precess throughout the simulation. The synthesis[21a] of $He@C_{20}H_{20}$ despite being thermodynamically unstable may encourage further studies on these systems including their possible synthesis.

**Acknowledgements**

PKC would like to thank DST, New Delhi for the J. C. Bose National Fellowship and the Indo-EU (HYPOMAP) project for financial assistance. MK and SP thank CSIR, New Delhi for their fellowships. We would like to thank Mr. Tamal Goswami and Mr. Sukanta Mondal for their help.

potential energy to improve substantially by atomic contraction at the expense of only a small increase in kinetic energy."

55. http://pubs.acs.org/cen/_interactive/explain-bond/index.html

**Figures**

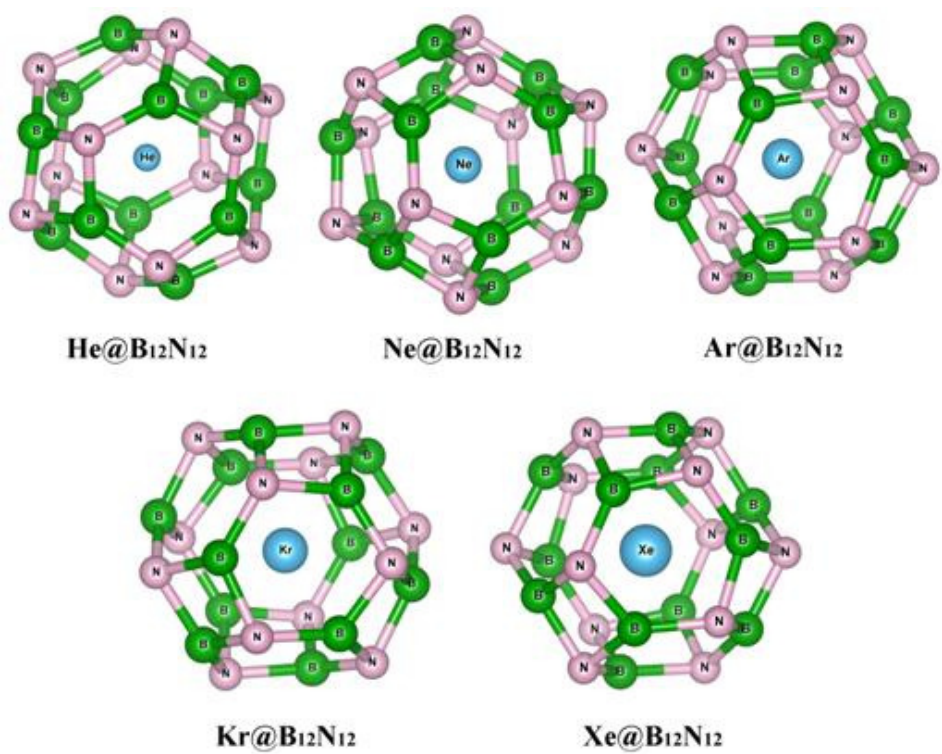

**Figure 1.** Optimized structures of Ng@$B_{12}N_{12}$ (Ng=He, Ne, Ar, Kr & Xe) at M05-2X/6-311+G(d,p) level for Ng= He, Ne, Ar, Kr and M05-2X/def2-TZVP level for Ng=Xe. (The Ng-N/B bonds are not shown for clarity of the structures).

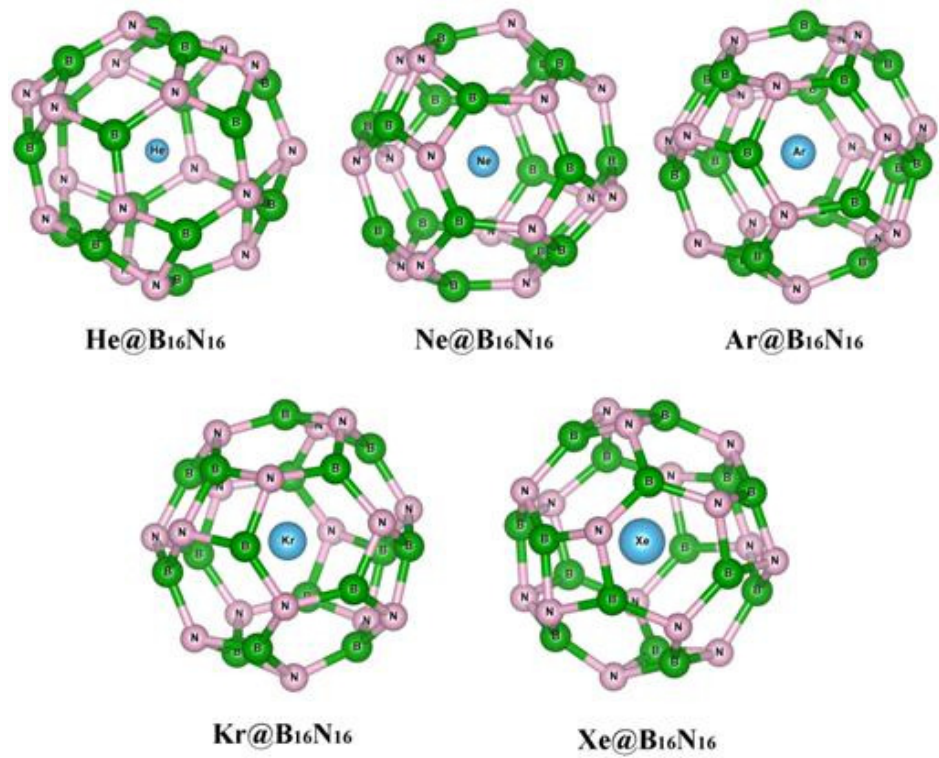

**Figure 2.** Optimized structures of Ng@B$_{16}$N$_{16}$ (Ng=He, Ne, Ar, Kr & Xe) at M05-2X/6-311+G(d,p) level for Ng= He, Ne, Ar, Kr and M05-2X/def2-TZVP level for Ng=Xe. (The Ng-N/B bonds are not shown for clarity of the structures).

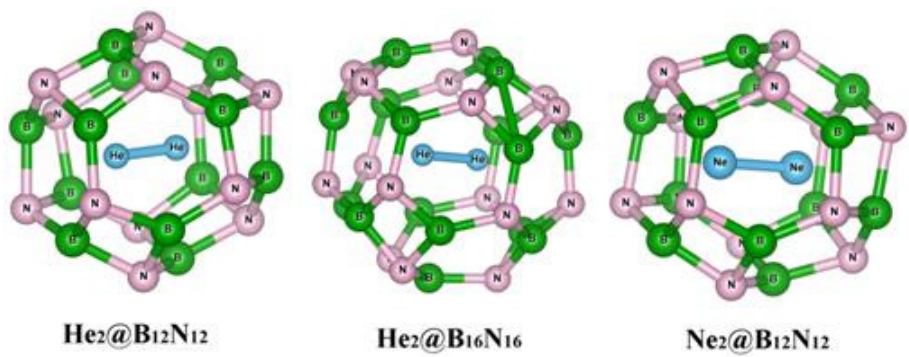

**Figure 3.** Optimized structures of $Ng_2@B_{12}N_{12}$ (Ng=He, Ne) and $He_2@B_{16}N_{16}$ at M05-2X/6-311+G(d,p) level. (The Ng-N/B bonds are not shown for clarity of the structures).

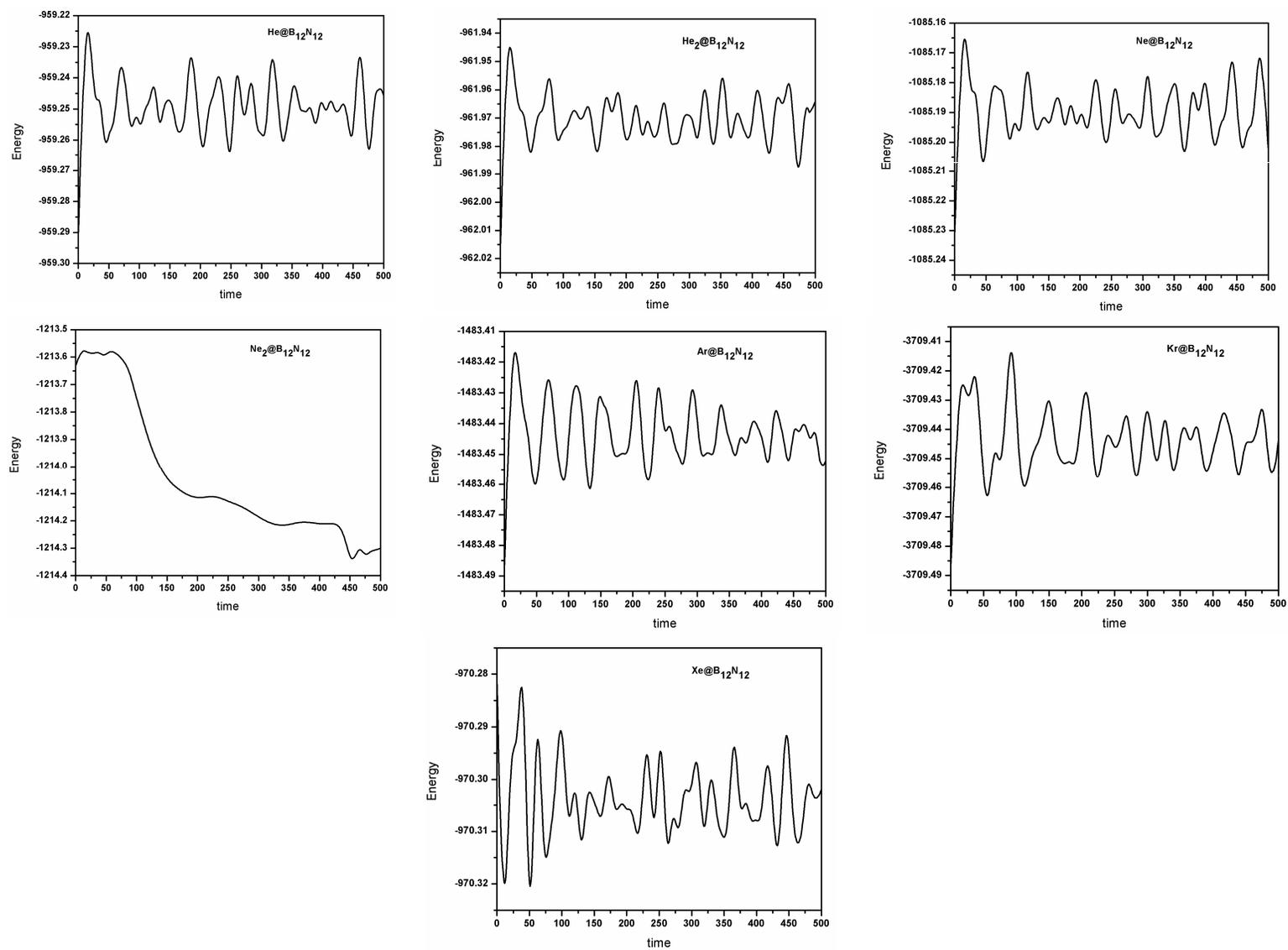

**Figure 4.** Time (in fs) evolution of energy (in au) for $Ng_n@B_{12}N_{12}$ systems.

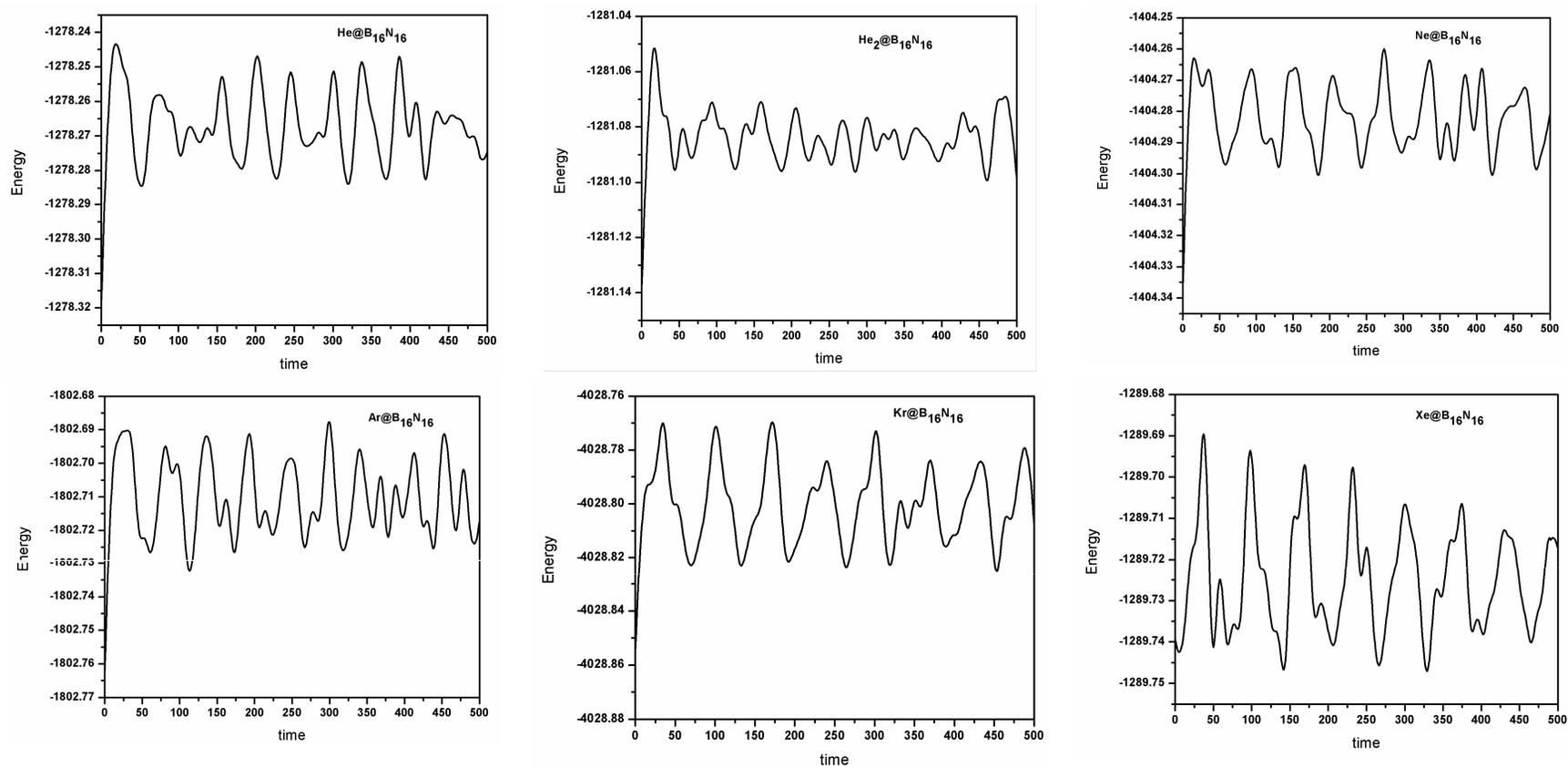

**Figure 5.** Time (in fs) evolution of energy (in au) for $Ng_n@B_{16}N_{16}$ systems.

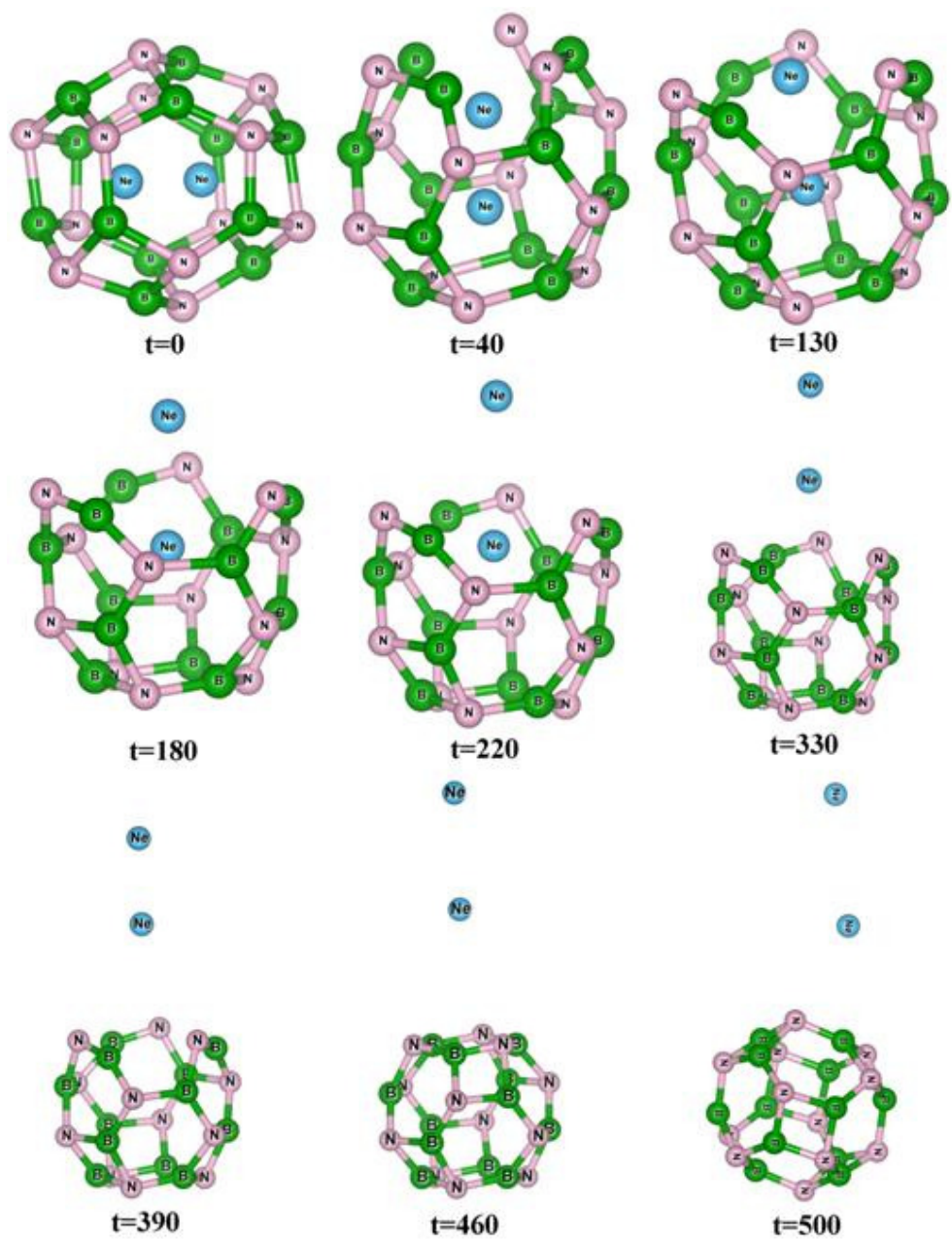

**Figure 6.** Structures of Ne$_2$@B$_{12}$N$_{12}$ systems at different time steps during simulation. (t in fs)

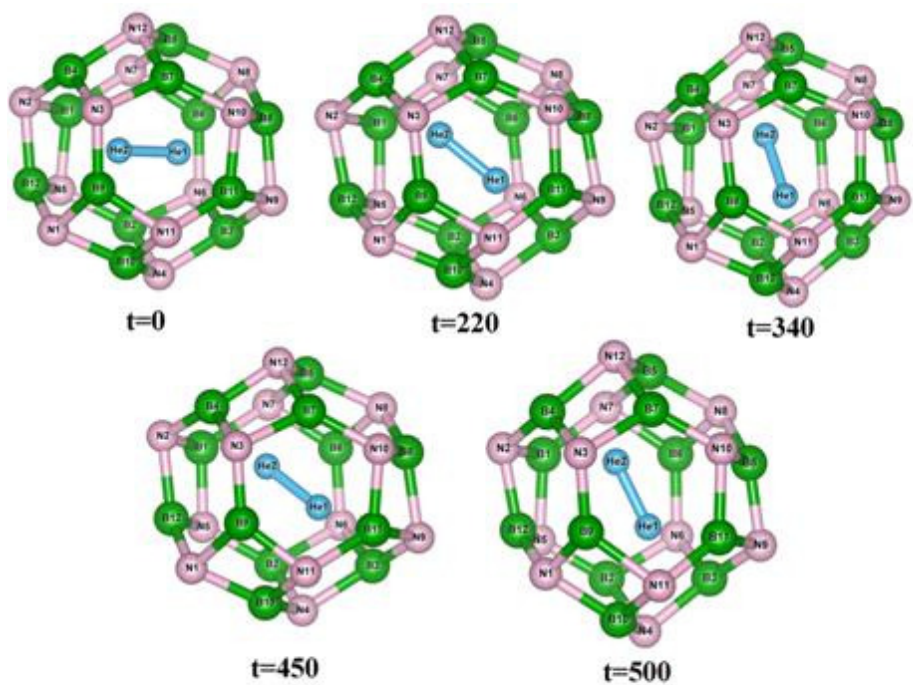

**Figure 7.** Snapshots of He$_2$@B$_{12}$N$_{12}$ system at different time steps to show the precession of He$_2$ unit. (t in fs)

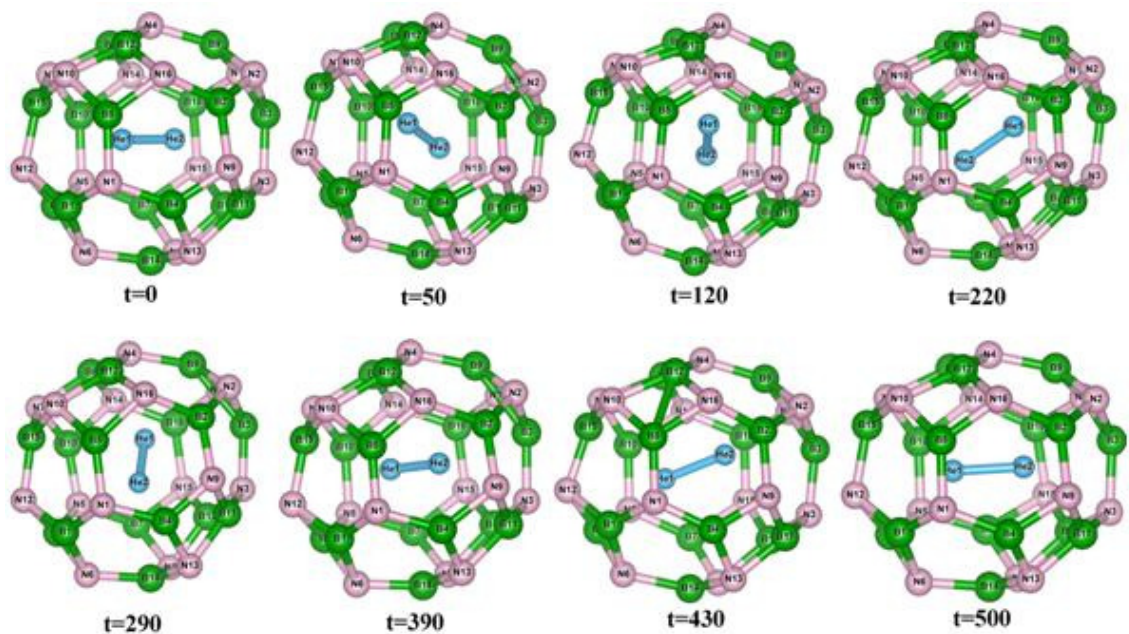

**Figure 8.** Snapshots of He$_2$@B$_{16}$N$_{16}$ system at different time steps to show the precession of He$_2$ unit. (t in fs)

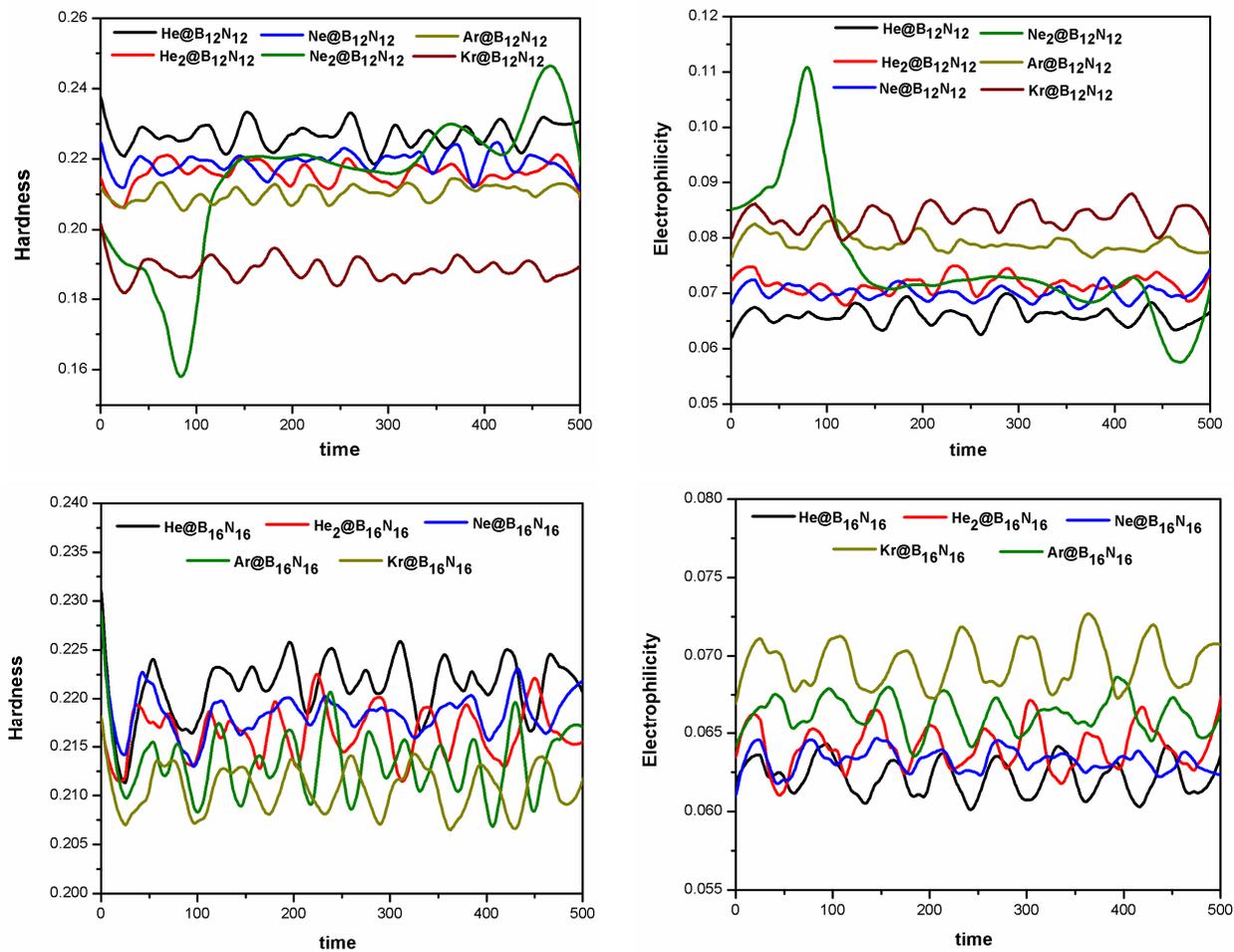

**Figure 9.** Time evolution (fs) of hardness (au) and electrophilicity (au) of $Ng_n@B_{12}N_{12}$ and $Ng_n@B_{16}N_{16}$ systems (Ng=He, Ne, Ar, Kr).

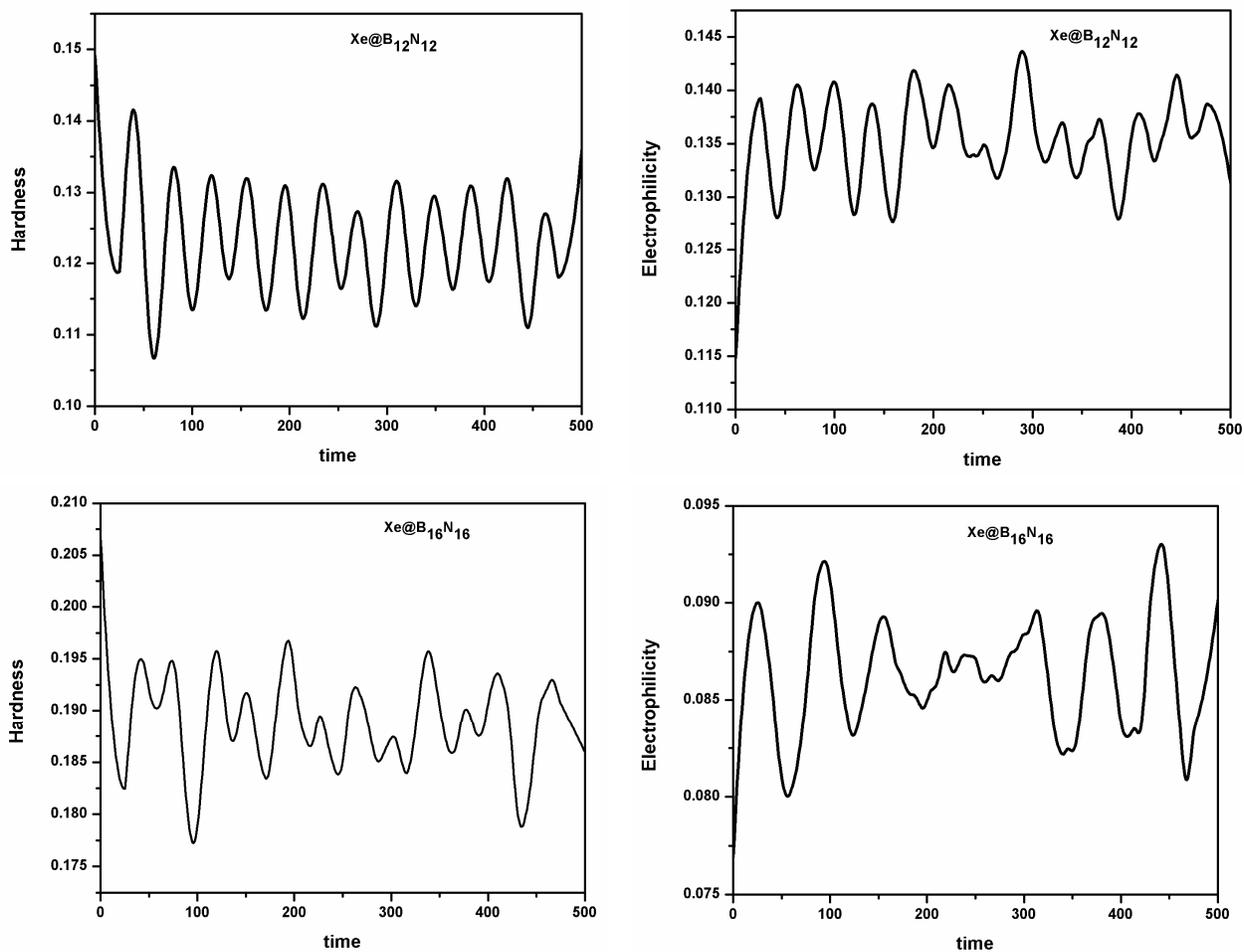

**Figure 10.** Time evolution (fs) of hardness (au) and electrophilicity (au) of Xe@B$_{12}$N$_{12}$ and Xe@B$_{16}$N$_{16}$ systems.

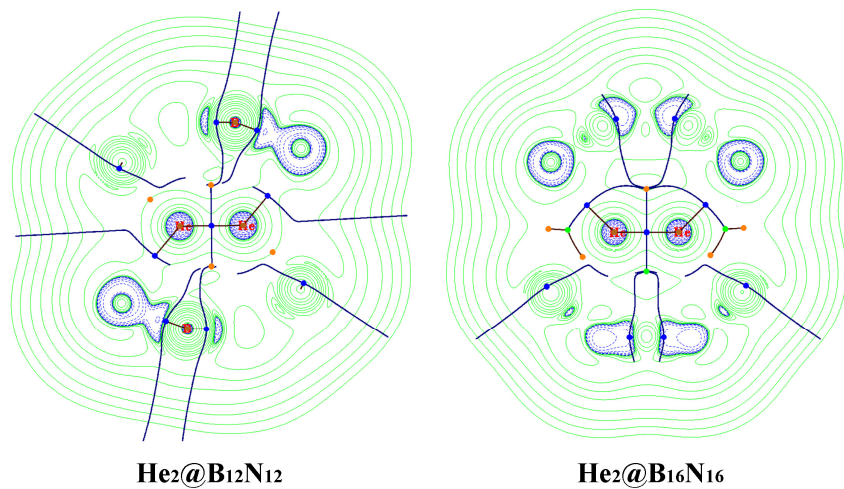

**Figure 11.** Contour line diagrams of the Laplacian of the electron density of He$_2$@B$_{12}$N$_{12}$, and He$_2$@B$_{16}$N$_{16}$ systems.

# Tables

**Table 1.** The ZPE and BSSE uncorrected dissociation energy ($D_e$, kcal/mol), ZPE corrected dissociation energy ($D_0$, kcal/mol), BSSE corrected dissociation energy ($D_{BSSE}$, kcal/mol), both ZPE and BSSE corrected dissociation energy ($D_0^{BSSE}$, kcal/mol) for the dissociation process: $Ng_n@B_{12}N_{12}/Ng_n@B_{16}N_{16} \rightarrow nNg + B_{12}N_{12}/B_{16}N_{16}$; reaction enthalpy ($\Delta H$, kcal/mol), and free energy change ($\Delta G$, kcal/mol) at 298K for the process: $nNg + B_{12}N_{12}/B_{16}N_{16} \rightarrow Ng_n@B_{12}N_{12}/Ng_n@B_{16}N_{16}$; hardness ($\eta$, eV), electrophilicity ($\omega$, eV) and polarizability ($\alpha$, au) of $B_{12}N_{12}/B_{16}N_{16}$ and $Ng_n@B_{12}N_{12}/Ng_n@B_{16}N_{16}$ structures calculated at M05-2X/6-311+G(d,p) level. For the systems having Xe atoms, the level is M05-2X/def2-TZVP.

| Systems | $D_e$ | $D_{BSSE}$ | $D_0$ | $D_0^{BSSE}$ | $\Delta H$ | $\Delta G$ | $\eta$ | $\omega$ | $\alpha$ |
|---|---|---|---|---|---|---|---|---|---|
| $B_{12}N_{12}$ | - | - | - | - | - | - | 9.681 (9.690) | 0.871 (0.851) | 154.956 (155.590) |
| $He@B_{12}N_{12}$ | -31.8 | -32.4 | -34.3 | -34.9 | 33.0 | 41.8 | 9.520 | 0.884 | 156.861 |
| $Ne@B_{12}N_{12}$ | -92.7 | -96.6 | -93.7 | -97.6 | 92.5 | 102.6 | 9.296 | 0.909 | 159.960 |
| $Ar@B_{12}N_{12}$ | -306.9 | -308.7 | -303.4 | -305.2 | 302.4 | 312.8 | 8.971 | 0.988 | 173.391 |
| $Kr@B_{12}N_{12}$ | -436.7 | -439.4 | -429.3 | -431.9 | 428.9 | 469.3 | 8.312 | 0.931 | 183.884 |
| $Xe@B_{12}N_{12}$ | -634.7 | -635.3 | -619.5 | -620.1 | 621.0 | 627.8 | 6.433 | 0.717 | 200.739 |
| $B_{16}N_{16}$ | - | - | - | - | - | - | 8.783 (8.831) | 0.911 (0.887) | 206.400 (206.884) |
| $He@B_{16}N_{16}$ | -6.9 | -7.3 | -8.8 | -9.2 | 7.8 | 16.2 | 9.128 | 0.845 | 207.072 |
| $Ne@B_{16}N_{16}$ | -22.7 | -25.8 | -23.9 | -27.0 | 22.9 | 32.5 | 9.029 | 0.851 | 208.333 |
| $Ar@B_{16}N_{16}$ | -113.9 | -115.5 | -113.4 | -115.0 | 112.3 | 122.8 | 8.870 | 0.868 | 217.898 |
| $Kr@B_{16}N_{16}$ | -182.3 | -183.7 | -180.2 | -181.6 | 179.2 | 190.1 | 8.792 | 0.891 | 224.091 |
| $Xe@B_{16}N_{16}$ | -298.0 | -298.6 | -292.8 | -293.4 | 292.1 | 302.9 | 8.393 | 0.854 | 236.465 |
| $He_2@B_{12}N_{12}$ | -156.5 | -157.5 | -162.2 | -163.2 | 159.7 | 176.8 | 9.011 | 0.904 | 161.485 |
| $Ne_2@B_{12}N_{12}$ | -451.1 | -458.4 | -448.3 | -455.6 | 447.1 | 464.0 | 8.320 | 0.987 | 172.560 |
| $He_2@B_{16}N_{16}$ | -67.0 | -68.0 | -71.9 | -72.9 | 69.7 | 86.3 | 8.976 | 0.850 | 209.902 |

**Table 2.** The NPA charge calculated at each center ($Q_k$, au) and maximum B-B and N-N distances ($r_{B-B}^{Max}$/ $r_{N-N}^{Max}$, Å) of the studied cage systems obtained at M05-2X/6-311+G(d,p) level. For the systems having Xe atoms, the level is M05-2X/def2-TZVP.

| Cages | Center | $Q_k$ (au) | $r_{B-B}^{Max}$/ $r_{N-N}^{Max}$ | Cages | Center | $Q_k$ (au) | $r_{B-B}^{Max}$/ $r_{N-N}^{Max}$ |
|---|---|---|---|---|---|---|---|
| $B_{12}N_{12}$ | B | +1.181, +1.182 | | $B_{16}N_{16}$ | B | +1.166, +1.201 | |
| | | +1.096, +1.097[a] | 4.368/4.808 | | | +1.089, +1.114[a] | 4.908/5.367 |
| | N | −1.181, −1.182 | 4.370/4.803[a] | | N | −1.160, −1.218 | 4.907/5.361[a] |
| | | −1.096, −1.097[a] | | | | −1.080, −1.142[a] | |
| He@$B_{12}N_{12}$ | B | +1.204 | 4.402/4.818 | He@$B_{16}N_{16}$ | B | +1.180 to +1.210 | 4.916/5.366 |
| | N | −1.212 | | | N | −1.178 to −1.237 | |
| | He | +0.089 | | | He | +0.058 | |
| Ne@$B_{12}N_{12}$ | B | +1.221, +1.222 | 4.456/4.839 | Ne@$B_{16}N_{16}$ | B | +1.190 to +1.213 | 4.932/5.369 |
| | N | −1.230, −1.231 | | | N | −1.186 to −1.243 | |
| | Ne | +0.107 | | | Ne | +0.08 | |
| Ar@$B_{12}N_{12}$ | B | +1.253, +1.254 | 4.562/4.930 | Ar@$B_{16}N_{16}$ | B | +1.224 to +1.236 | 4.990/5.391 |
| | N | −1.279, −1.280 | | | N | −1.230 to −1.284 | |
| | Ar | +0.317 | | | Ar | +0.256 | |
| Kr@$B_{12}N_{12}$ | B | +1.255 | 4.609/4.994 | Kr@$B_{16}N_{16}$ | B | +1.238 to +1.244 | 5.025/5.411 |
| | N | −1.295 | | | N | −1.250 to −1.301 | |
| | Kr | +0.486 | | | Kr | +0.377 | |
| Xe@$B_{12}N_{12}$ | B | +1.086 to +1.088 | 4.672/5.099 | Xe@$B_{16}N_{16}$ | B | +1.129 to +1.131 | 5.069/5.447 |
| | N | −1.151, −1.152 | | | N | −1.149 to −1.191 | |
| | Xe | +0.775 | | | Xe | +0.458 | |
| $He_2$@$B_{12}N_{12}$ | B | +1.216 to +1.234 | 4.554/4.961 | $He_2$@$B_{16}N_{16}$ | B | +1.179 to +1.235 | 5.022/5.400 |
| | N | −1.237 to −1.239 | | | N | −1.190 to −1.257 | |
| | He | +0.085 | | | He | +0.066, +0.067 | |

[a] the values are at M05-2X/def2-TZVP level.

**Table 3.** The highest occupied atomic orbital (HOAO) energy of He atom ($E_{HOAO}$), HOMO energy of [$He_2$] having bond length as those in $He_2@B_{12}N_{12}$ and $He_2@B_{16}N_{16}$, and MO energy involving the He-He interaction in $He_2@B_{12}N_{12}$. All the energy terms and their differences are calculated in eV unit and at M05-2X/6-311+G(d,p) level.

| Ng | $E_{HOAO}$ He | $E_{HOMO}$ [$He_2$] In $B_{12}N_{12}$ | In $B_{16}N_{16}$ | $\Delta E_{HOMO}$ [$He_2$] – He In $B_{12}N_{12}$ | In $B_{16}N_{16}$ | E ($He_2$) in $He_2@B_{12}N_{12}$ | $\Delta E$ ($He_2$) in $He_2@B_{12}N_{12}$ – [$He_2$] |
|---|---|---|---|---|---|---|---|
| He | -20.768 | -17.814 | -18.674 | 2.954 | 2.094 | -17.695 | 0.119 |

**Table 4.** The valence orbital population of Ng centers in $Ng_n@B_{12}N_{12}$ and $Ng_n@B_{16}N_{16}$ systems.

| Systems | Ng center | Valence Orbital Population |
|---|---|---|
| $He@B_{12}N_{12}$ | He | $1s^{1.910}$ |
| $He_2@B_{12}N_{12}$ | He, He | $1s^{1.912}$ |
| $Ne@B_{12}N_{12}$ | Ne | $2s^{1.941}\ 2p_x^{1.983}\ 2p_y^{1.983}\ 2p_z^{1.983}$ |
| $Ne_2@B_{12}N_{12}$ | Ne, Ne | $2s^{1.937}\ 2p_x^{1.983}\ 2p_y^{1.978}\ 2p_z^{1.977}$ |
| $Ar@B_{12}N_{12}$ | Ar | $3s^{1.850}\ 3p_x^{1.931}\ 3p_y^{1.932}\ 3p_z^{1.931}$ |
| $Kr@B_{12}N_{12}$ | Kr | $4s^{1.807}\ 4p_x^{1.883}\ 4p_y^{1.883}\ 4p_z^{1.883}$ |
| $Xe@B_{12}N_{12}$ | Xe | $5s^{1.818}\ 5p_x^{1.787}\ 5p_y^{1.786}\ 5p_z^{1.786}$ |
| $He@B_{16}N_{16}$ | He | $1s^{1.941}$ |
| $He_2@B_{16}N_{16}$ | He, He | $1s^{1.932}$ |
| $Ne@B_{16}N_{16}$ | Ne | $2s^{1.963}\ 2p_x^{1.985}\ 2p_y^{1.985}$ |
| $Ar@B_{16}N_{16}$ | Ar | $3s^{1.899}\ 3p_x^{1.942}\ 3p_y^{1.942}\ 3p_z^{1.942}$ |
| $Kr@B_{16}N_{16}$ | Kr | $4s^{1.869}\ 4p_x^{1.909}\ 4p_y^{1.909}\ 4p_z^{1.909}$ |
| $Xe@B_{16}N_{16}$ | Xe | $5s^{1.891}\ 5p_x^{1.879}\ 5p_y^{1.879}\ 5p_z^{1.879}$ |

**Table 5.** Electron density descriptors (au) at the bond critical points (BCP) in between Ng and N/B/Ng/Be atoms obtained from the wave functions generated at M05-2X/6-311+G(d,p) level. For the systems having Xe atoms, the level is M05-2X/def2-TZVP.

| Systems | BCP | $\rho(r_c)$ | $\nabla^2\rho(r_c)$ | $G(r_c)$ | $V(r_c)$ | $H(r_c)$ | $-G(r_c)/V(r_c)$ | $G(r_c)/\rho(r_c)$ | Class |
|---|---|---|---|---|---|---|---|---|---|
| He@$B_{12}N_{12}$ | He-N | 0.0223 | 0.1052 | 0.0237 | -0.0211 | 0.0026 | 1.1232 | 1.0628 | $W^n$ |
| Ne@$B_{12}N_{12}$ | Ne-N | 0.0311 | 0.1645 | 0.0381 | -0.0351 | 0.0030 | 1.0855 | 1.2251 | $W^n$ |
| Ar@$B_{12}N_{12}$ | Ar-N | 0.0475 | 0.2201 | 0.0541 | -0.0532 | 0.0009 | 1.0169 | 1.1389 | $W^n$ |
| Kr@$B_{12}N_{12}$ | Kr-N | 0.0544 | 0.2195 | 0.0582 | -0.0616 | -0.0034 | 0.9448 | 1.0699 | $W^c$ |
|  | Kr-B | 0.0507 | 0.1381 | 0.0484 | -0.0623 | -0.0139 | 0.7769 | 0.9546 | C |
| Xe@$B_{12}N_{12}$ | Xe-N | 0.0625 | 0.1904 | 0.0578 | -0.0681 | -0.0103 | 0.8488 | 0.9248 | C |
|  | Xe-B | 0.0572 | 0.1256 | 0.0495 | -0.0677 | -0.0182 | 0.7312 | 0.8654 | C |
| He@$B_{16}N_{16}$ | He-N | 0.0122 | 0.0551 | 0.0110 | -0.0082 | 0.0028 | 1.3415 | 0.9016 | D |
|  | He-B | 0.0111 | 0.0492 | 0.0104 | -0.0084 | 0.0020 | 1.2381 | 0.9369 | D |
| Ne@$B_{16}N_{16}$ | Ne-N | 0.0167 | 0.0750 | 0.0179 | -0.0171 | 0.0008 | 1.0468 | 1.0719 | $W^n$ |
|  | Ne-B | 0.0152 | 0.0635 | 0.0157 | -0.0155 | 0.0002 | 1.0129 | 1.0329 | $W^n$ |
| Ar@$B_{16}N_{16}$ | Ar-N | 0.0331 | 0.1405 | 0.0331 | -0.0311 | 0.0020 | 1.0643 | 1.0000 | $W^n$ |
|  | Ar-B | 0.0252 | 0.0879 | 0.0232 | -0.0244 | -0.0012 | 0.9508 | 0.9206 | C |
| Kr@$B_{16}N_{16}$ | Kr-N | 0.0316 | 0.1152 | 0.0288 | -0.0287 | 0.0001 | 1.0035 | 0.9114 | D |
|  | Kr-B | 0.0302 | 0.0904 | 0.0259 | -0.0291 | -0.0032 | 0.8900 | 0.8576 | C |
| Xe@$B_{16}N_{16}$ | Xe-N | 0.0389 | 0.1161 | 0.0320 | -0.0350 | -0.0030 | 0.9143 | 0.8226 | C |
|  | Xe-B | 0.0348 | 0.0993 | 0.0285 | -0.0321 | -0.0036 | 0.8879 | 0.8190 | C |
| $He_2$@$B_{12}N_{12}$ | He-He | 0.0873 | 0.5167 | 0.1383 | -0.1475 | -0.0092 | 0.9376 | 1.5842 | $W^c$ |
|  | He-N | 0.0477 | 0.2455 | 0.0595 | -0.0577 | 0.0018 | 1.0312 | 1.2474 | $W^n$ |
| $He_2$@$B_{16}N_{16}$ | He-He | 0.0559 | 0.3304 | 0.0826 | -0.0826 | 0.0000 | 1.0000 | 1.4776 | $W^n$ |
|  | He-N | 0.0311 | 0.1585 | 0.0353 | -0.0309 | 0.0044 | 1.1424 | 1.1350 | $W^n$ |

| | | | | | | | | |
|---|---|---|---|---|---|---|---|---|
| Kr@BeCN$_2$ | Kr-Be | 0.0365 | 0.1628 | 0.0450 | -0.0493 | -0.0043 | 0.9128 | 1.2329 | W$^c$ |
| Kr@BeO | Kr-Be | 0.0327 | 0.1542 | 0.0411 | -0.0437 | -0.0026 | 0.9405 | 1.2569 | W$^c$ |
| Xe@BeCN$_2$ | Xe-Be | 0.0332 | 0.1137 | 0.0342 | -0.0399 | -0.0057 | 0.8571 | 1.0301 | W$^c$ |
| Xe@BeO | Xe-Be | 0.0290 | 0.1117 | 0.0312 | -0.0345 | -0.0033 | 0.9043 | 1.0759 | W$^c$ |

**Table 6.** EDA results of the Ng$_n$@B$_{12}$N$_{12}$ and Ng$_n$@B$_{16}$N$_{16}$ systems studied at BP86-D/TZ2P//M05-2X/6-311+G(d,p) level. All energy terms are in kcal/mol. (For Xe cases, the level is BP86-D/TZ2P//M05-2X/def2-TZVP)

| Systems | Fragments | $\Delta E_{int}$ | $\Delta E_{pauli}$ | $\Delta E_{elstat}$ | $\Delta E_{orb}$ | $\Delta E_{disp}$ |
|---|---|---|---|---|---|---|
| He@B$_{12}$N$_{12}$ | [He]+[B$_{12}$N$_{12}$] | 33.0 | 72.4 | -21.7 | -14.2 | -3.5 |
| Ne@B$_{12}$N$_{12}$ | [Ne]+[B$_{12}$N$_{12}$] | 96.5 | 195.6 | -73.9 | -22.3 | -3.0 |
| Ar@B$_{12}$N$_{12}$ | [Ar]+[B$_{12}$N$_{12}$] | 252.8 | 616.8 | -252.3 | -110.3 | -1.4 |
| Kr@B$_{12}$N$_{12}$ | [Kr]+[B$_{12}$N$_{12}$] | 346.5 | 899.9 | -388.5 | -163.6 | -1.3 |
| Xe@B$_{12}$N$_{12}$ | [Xe]+[B$_{12}$N$_{12}$] | 483.8 | 1359.6 | -611.5 | -263.2 | -1.2 |
| He$_2$@B$_{12}$N$_{12}$ | [He$_2$]+[B$_{12}$N$_{12}$] | 113.2 | 238.9 | -73.4 | -46.0 | -6.3 |
| He$_2$ in B$_{12}$N$_{12}$ | [He]+[He] | 27.4 | 40.5 | -7.8 | -5.4 | 0.0 |
| He@B$_{16}$N$_{16}$ | [He]+[B$_{16}$N$_{16}$] | 8.1 | 29.1 | -8.3 | -5.5 | -7.2 |
| Ne@B$_{16}$N$_{16}$ | [Ne]+[B$_{16}$N$_{16}$] | 28.9 | 75.7 | -29.0 | -7.9 | -10.0 |
| Ar@B$_{16}$N$_{16}$ | [Ar]+[B$_{16}$N$_{16}$] | 109.2 | 277.4 | -115.5 | -46.9 | -5.9 |
| Kr@B$_{16}$N$_{16}$ | [Kr]+[B$_{16}$N$_{16}$] | 161.4 | 431.1 | -190.4 | -74.1 | -5.2 |
| Xe@B$_{16}$N$_{16}$ | [Xe]+[B$_{16}$N$_{16}$] | 254.0 | 707.5 | -328.0 | -121.4 | -4.13 |
| He$_2$@B$_{16}$N$_{16}$ | [He$_2$]+[B$_{16}$N$_{16}$] | 52.1 | 119.2 | -36.2 | -21.8 | -9.1 |
| He$_2$ in B$_{16}$N$_{16}$ | [He]+[He] | 15.0 | 21.7 | -4.2 | -2.5 | 0.0 |

# Does Confinement Force Marriage between Two Unwilling Partners?: A Case Study of He$_2$@B$_m$N$_m$ (m=12, 16)

Munmun Khatua, Sudip Pan and Pratim K. Chattaraj*

*Department of Chemistry and Center for Theoretical Studies,*

*Indian Institute of Technology, Kharagpur 721302, India.*

[*]*To whom correspondence should be addressed. E-mail: pkc@chem.iitkgp.ernet.in,*

*Telephone: +91 3222 283304, Fax: 91-3222-255303*

## Table of Contents

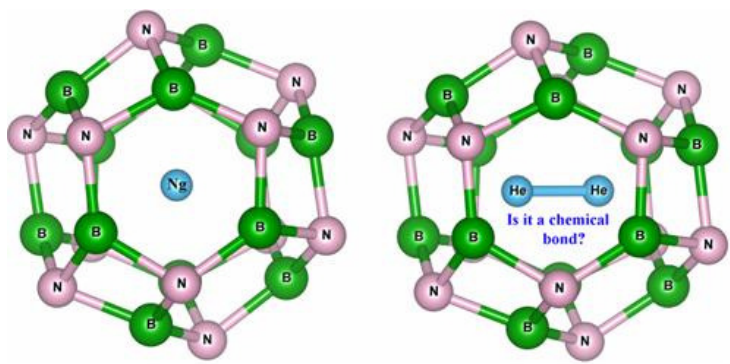

# Does Confinement Force Marriage between Two Unwilling Partners?: A Case Study of He$_2$@B$_m$N$_m$ (m=12, 16)

Munmun Khatua, Sudip Pan and Pratim K. Chattaraj*

*Department of Chemistry and Center for Theoretical Studies,*

*Indian Institute of Technology, Kharagpur 721302, India.*

*To whom correspondence should be addressed. E-mail: pkc@chem.iitkgp.ernet.in,

Telephone: +91 3222 283304, Fax: 91-3222-255303

## Supporting Information

**Table S1.** Interaction energies ($\Delta E_{int}$, kcal/mol) and preparation energies ($\Delta E_{prep}$, kcal/mol) of cages and He$_2$ dimer of the Ng$_n$@B$_{12}$N$_{12}$ and Ng$_n$@B$_{16}$N$_{16}$ systems calculated at M05-2X/6-311+G(d,p) level.

| Systems | $\Delta E_{int}$ | $\Delta E_{prep}$ (B$_{12}$N$_{12}$) | Systems | $\Delta E_{prep}$ (B$_{16}$N$_{16}$) | $\Delta E_{int}$ | $\Delta E_{prep}$ (He$_2$) In B$_{12}$N$_{12}$ | $\Delta E_{prep}$ (He$_2$) In B$_{16}$N$_{16}$ |
|---|---|---|---|---|---|---|---|
| He@B$_{12}$N$_{12}$ | 30.9 | 0.9 | He@B$_{16}$N$_{16}$ | 0.2 | 6.7 | - | - |
| Ne@B$_{12}$N$_{12}$ | 86.3 | 6.4 | Ne@B$_{16}$N$_{16}$ | 1.1 | 21.6 | - | - |
| Ar@B$_{12}$N$_{12}$ | 271.7 | 35.2 | Ar@B$_{16}$N$_{16}$ | 8.6 | 105.3 | - | - |
| Kr@B$_{12}$N$_{12}$ | 376.6 | 60.1 | Kr@B$_{16}$N$_{16}$ | 16.7 | 166.0 | - | - |
| Xe@B$_{12}$N$_{12}$ | 522.1 | 112.6 | Xe@B$_{16}$N$_{16}$ | 33.4 | 264.6 | - | - |
| He$_2$@B$_{12}$N$_{12}$ | 117.1 | 12.1 | He$_2$@B$_{16}$N$_{16}$ | 4.0 | 48.6 | 27.3 | 14.4 |

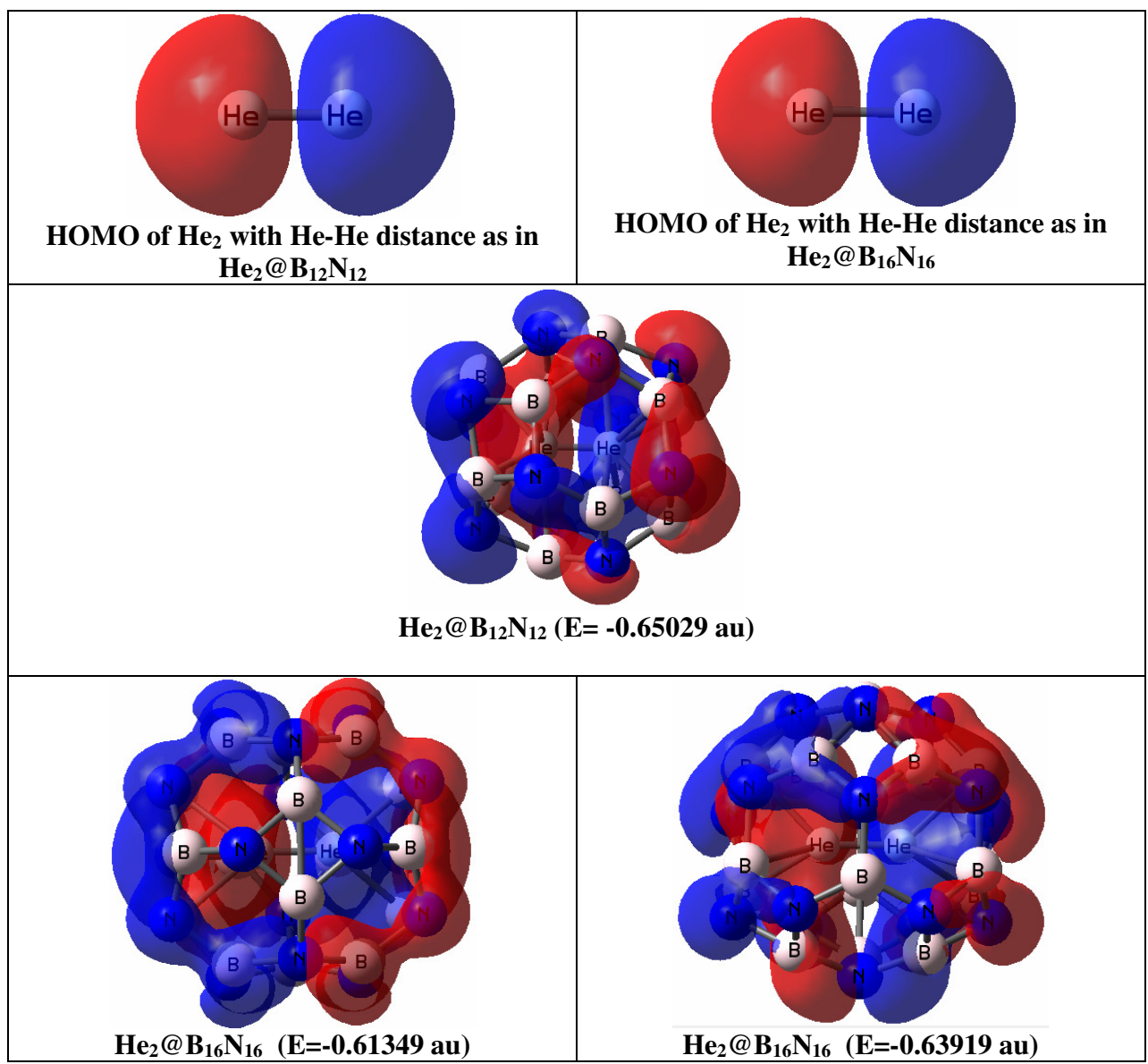

**Figure S1.** HOMO of free He$_2$ having bond distances as in He$_2$@B$_{12}$N$_{12}$ and He$_2$@B$_{16}$N$_{16}$ systems and related MOs of He$_2$@B$_{12}$N$_{12}$ and He$_2$@B$_{16}$N$_{16}$ showing He-He interaction.

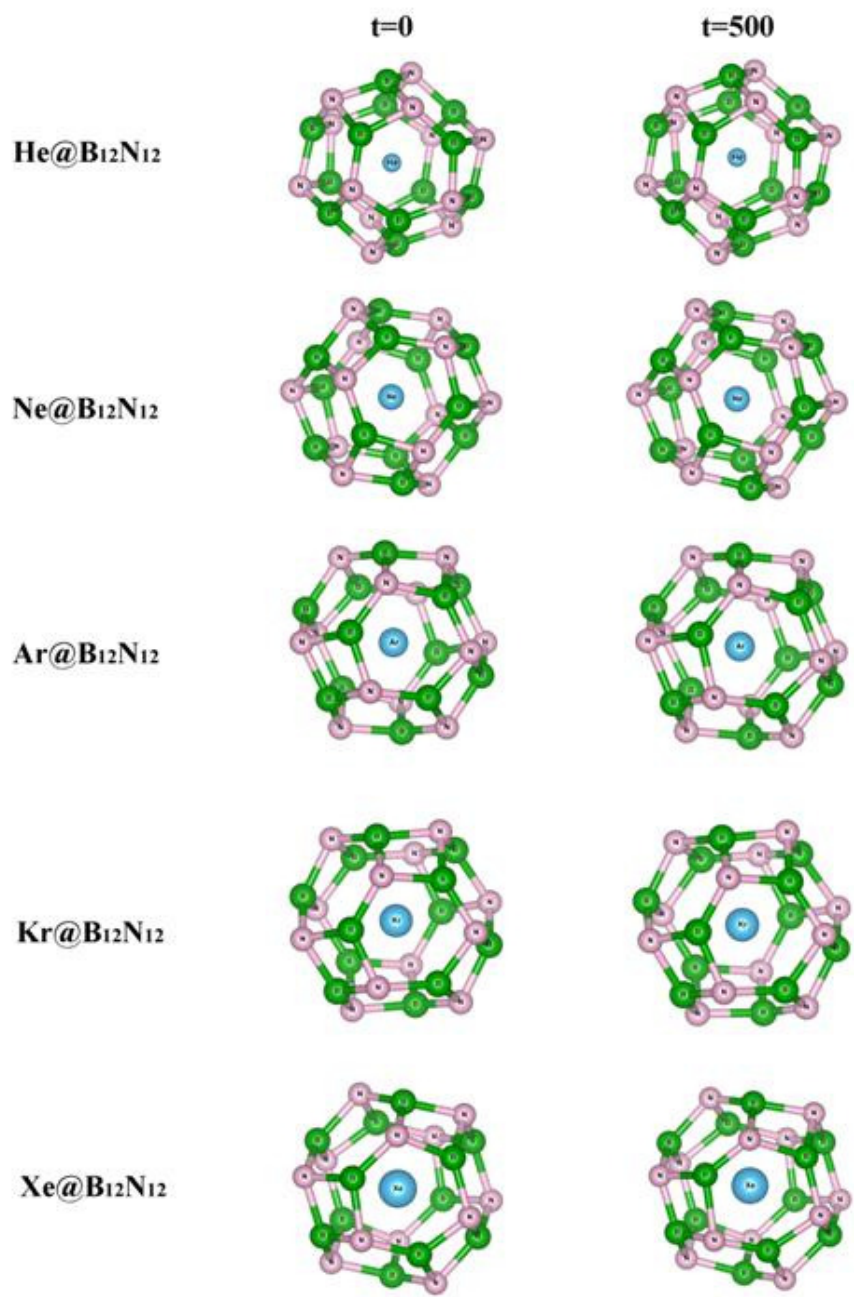

**Figure S2.** Structures of Ng@$B_{12}N_{12}$ systems at t=0 and 500 fs during simulation.

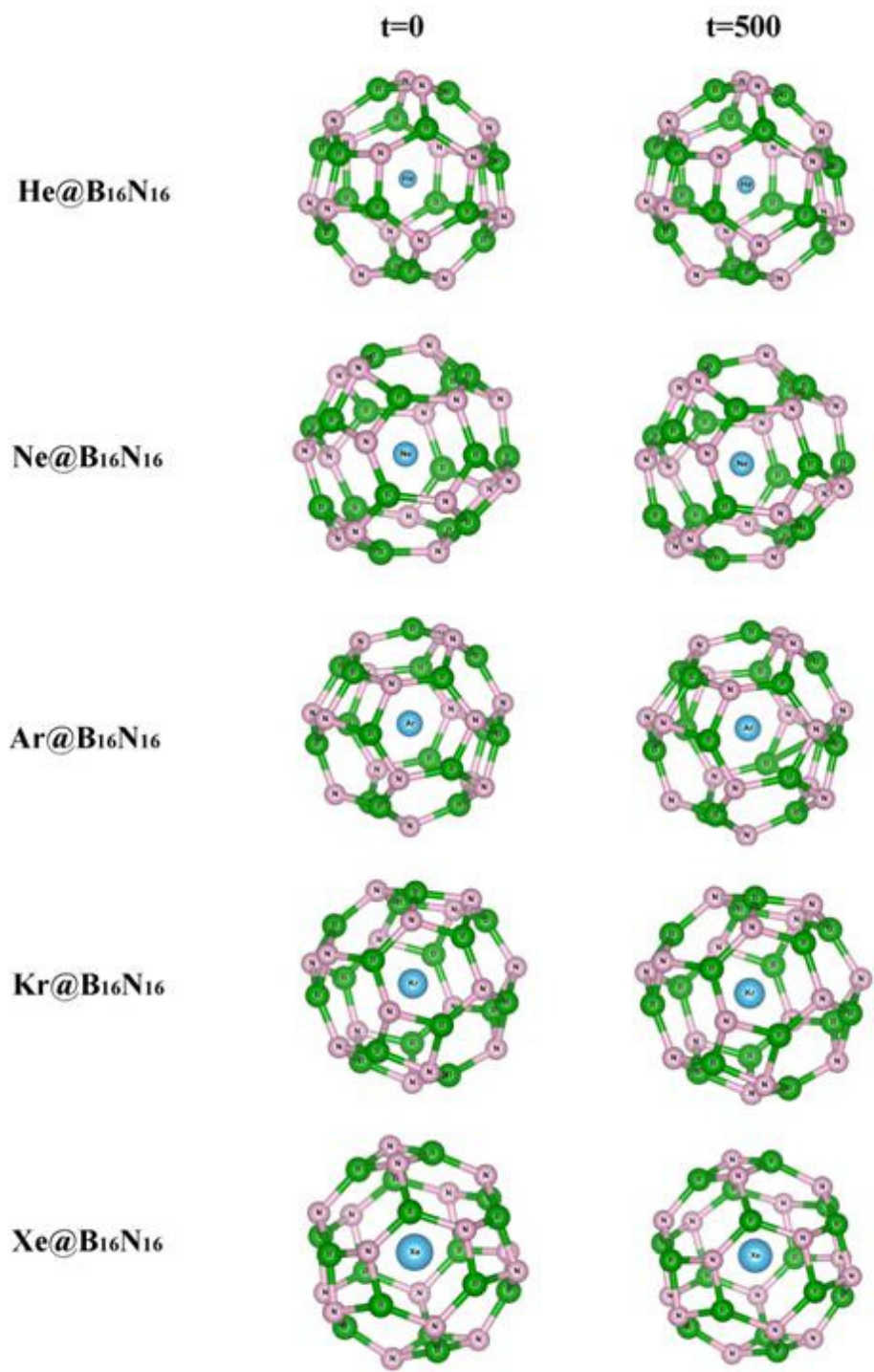

**Figure S3.** Structures of Ng@$B_{16}N_{16}$ systems at t=0 and 500 fs during simulation.

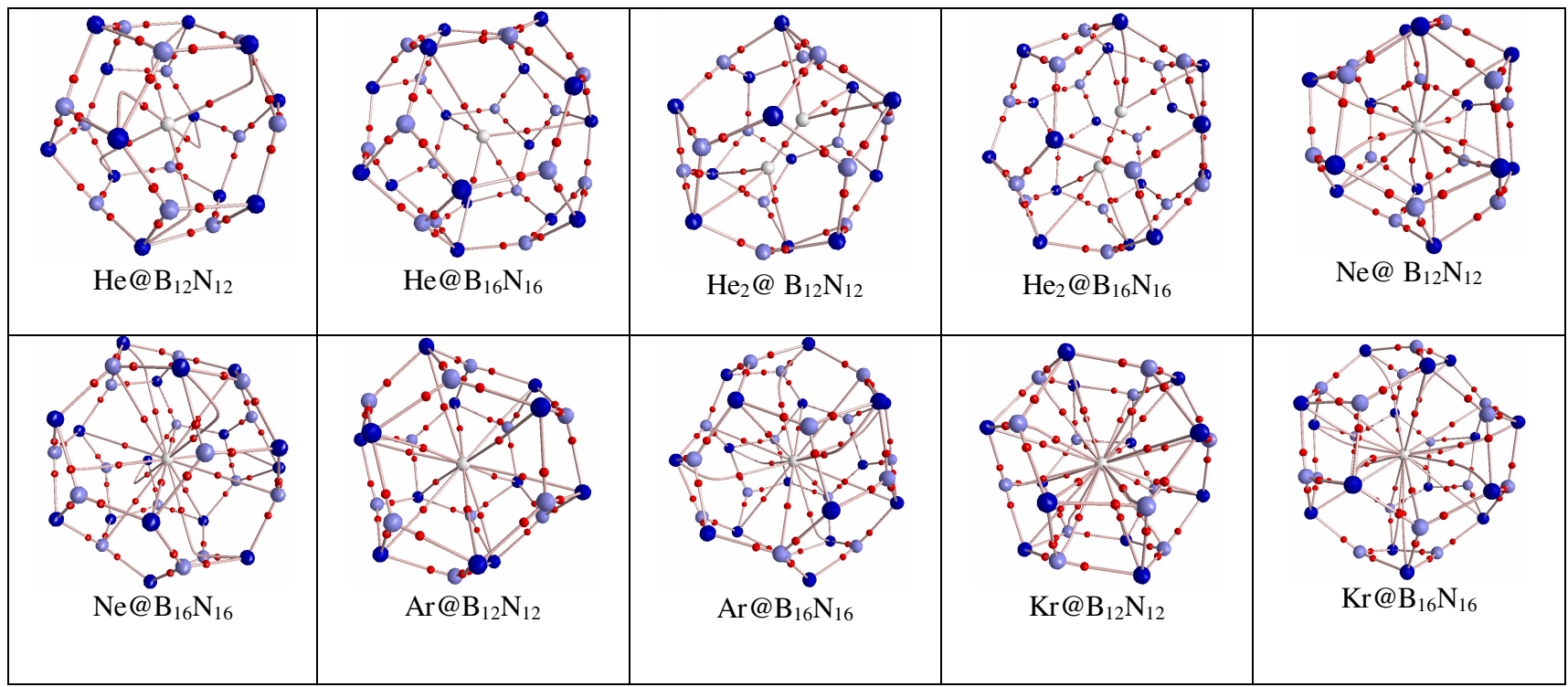

**Figure S4.** The molecular graphs of $Ng_n@B_{12}N_{12}$ and $Ng_n@B_{16}N_{16}$ systems.

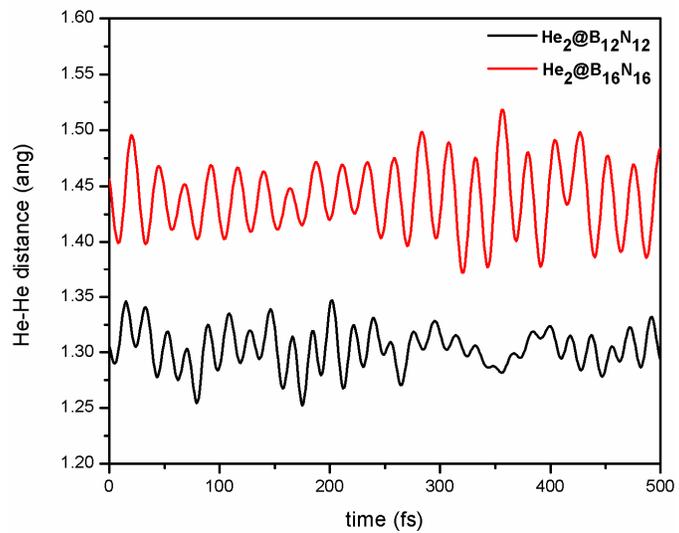

**Figure S5.** Plot of the He-He bond distance vs time for He$_2$@B$_{12}$N$_{12}$ and He$_2$@B$_{16}$N$_{16}$ systems.

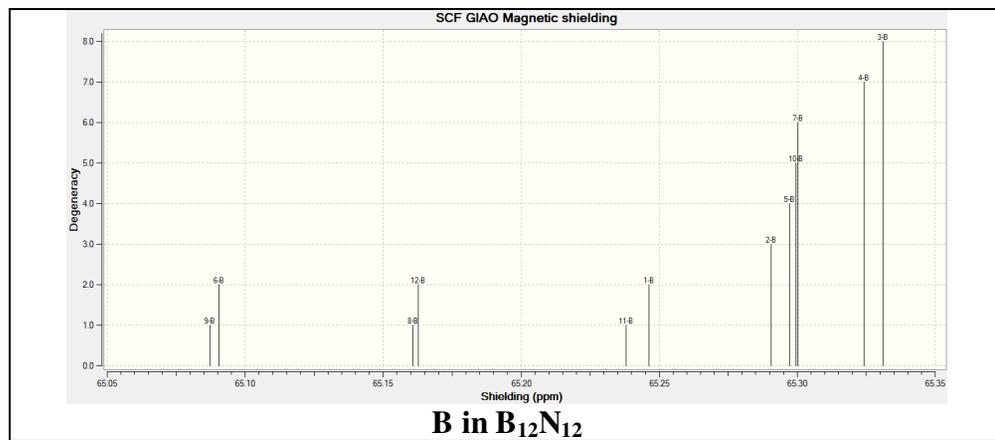
**B in $B_{12}N_{12}$**

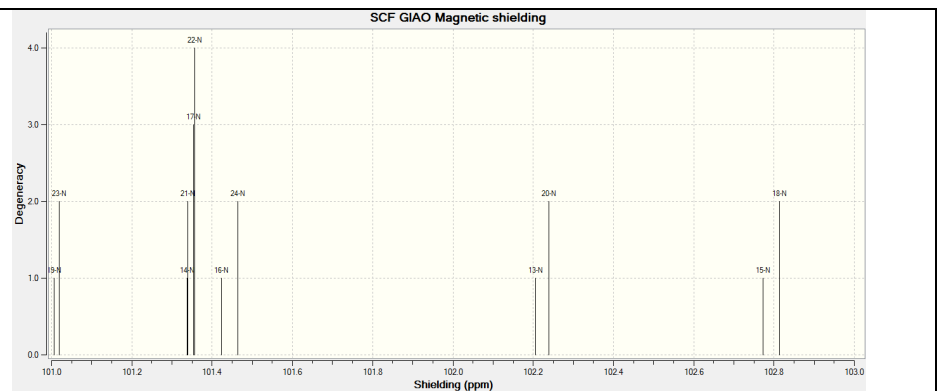
**N in $B_{12}N_{12}$**

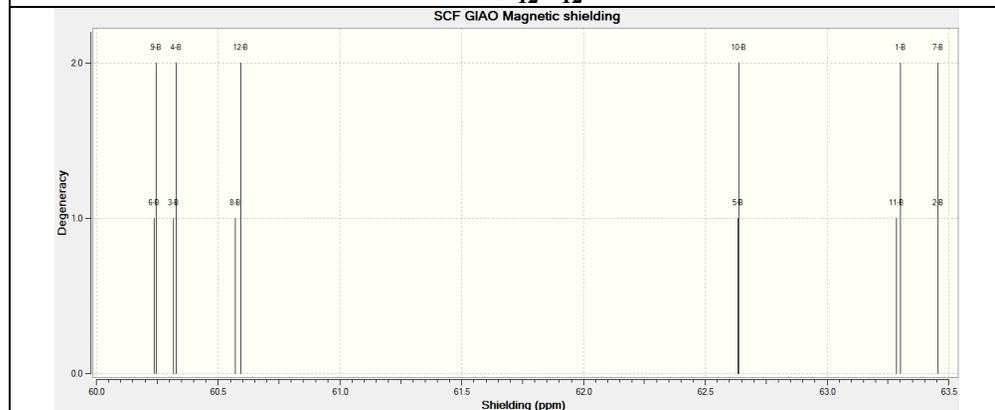
**B in $He_2@B_{12}N_{12}$**

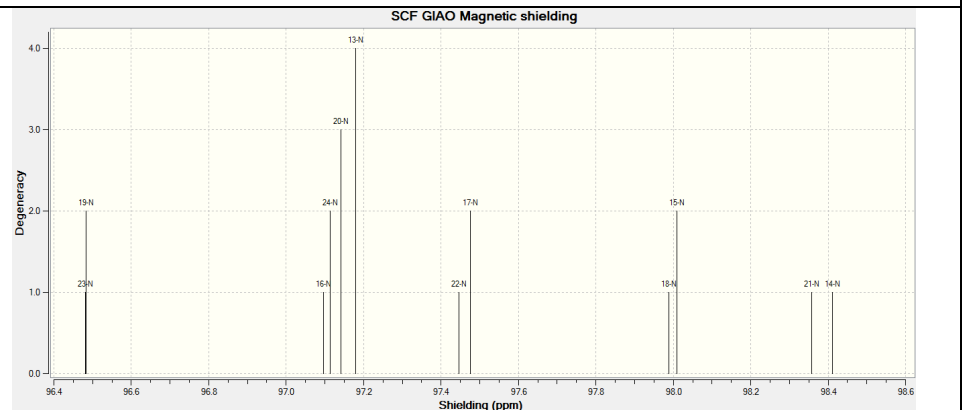
**N in $He_2@B_{12}N_{12}$**

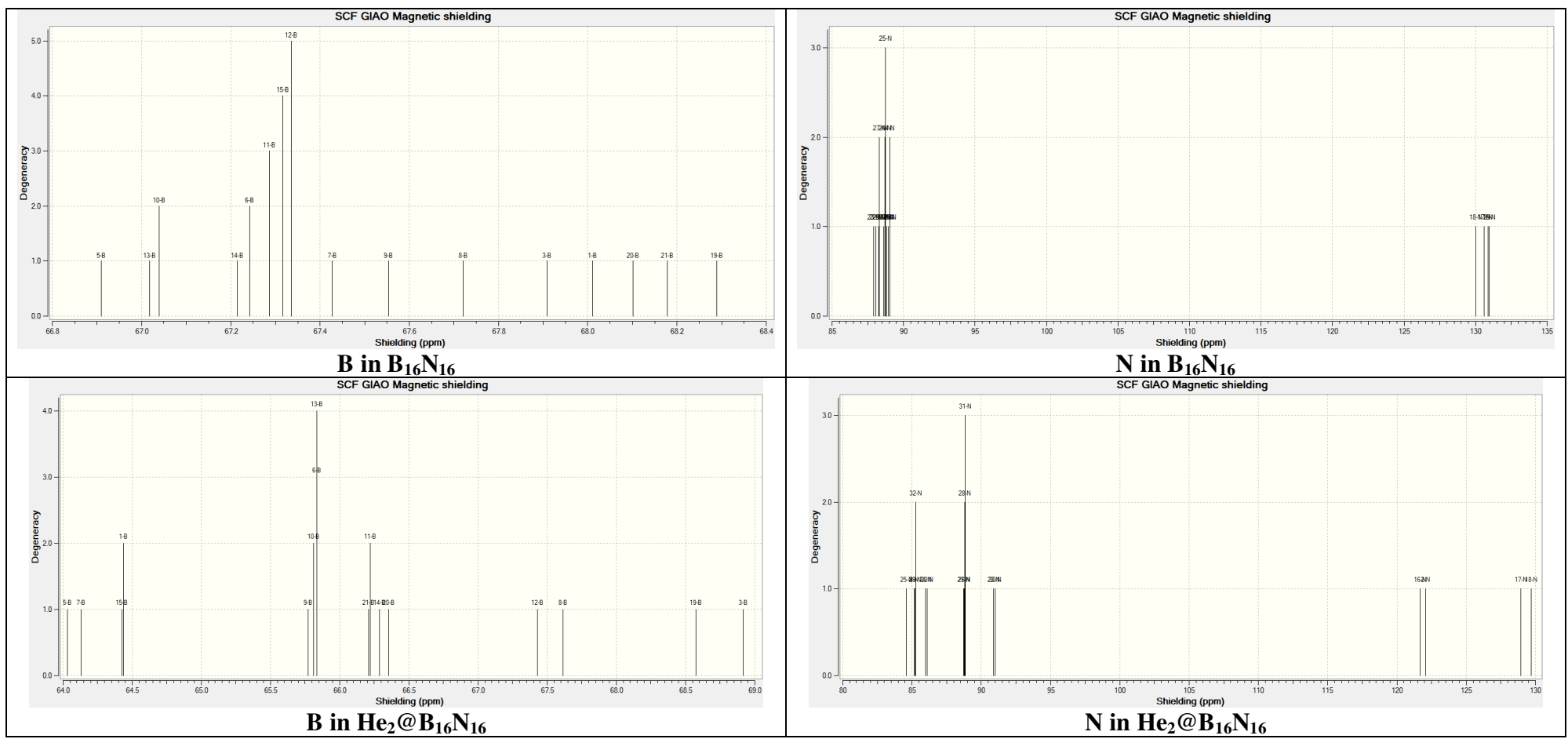

**Figure S6.** Magnetic shielding (in ppm) of B and N atoms in $B_{12}N_{12}$, $B_{16}N_{16}$, $He_2@B_{12}N_{12}$ and $He_2@B_{16}N_{16}$ systems.